# CRYSTAL CHEMICAL INSIGHTS ON LEAD IODIDE PEROVSKITES DOPING FROM REVISED EFFECTIVE RADII OF METAL IONS

*Ekaterina I. Marchenko [1,2], Sergey A. Fateev [1], Nikolay N. Eremin [2], Chen Qi [3], Eugene A. Goodilin [1,4], and Alexey B. Tarasov [1,4]**

[1] Laboratory of New Materials for Solar Energetics, Faculty of Materials Science, Lomonosov Moscow State University, 1 Lenin Hills, 119991, Moscow, Russia

[2] Department of Geology, Lomonosov Moscow State University, 1 Lenin Hills, 119991, Moscow, Russia

[3] Beijing Key Laboratory of Nanophotonics and Ultrafine Optoelectronic Systems, School of Materials Science & Engineering, Beijing Institute of Technology, Beijing 100081, P. R. China

[4] Department of Chemistry, Lomonosov Moscow State University, 1 Lenin Hills, 119991, Moscow, Russia

Corresponding Author: * Alexey B. Tarasov, e-mail: alexey.bor.tarasov@yandex.ru

**ABSTRACT.** Over the last few years of the heyday of hybrid halide perovskites, so many metal cations additives have been tested to improve their optoelectronic properties that it is already difficult to find an element that has not yet been tried. In general, the variety of these approaches is united under the name "doping", however, there is currently no clear understanding of the mechanisms of the influence of the metal ion additives on the properties of the lead halide perovskite materials. For many ions there is even no consensus on the most fundamental questions: what lattice position does a given ion occupy and is it incorporated in the structure at all? Here, we derived a system of effective radii of different metal ions in the iodine environment for the set of iodide compounds and reveal their crystal chemical role in the $APbI_3$ perovskites. We analysed the possible lattice positions for 40 most common monovalent, divalent, and trivalent metals to reveal whether they could successfully enter into the perovskite structures. We show that, at most, three parameters — effective size, electronegativity and the softness of metal ions are the main ones for crystal chemical analysis of the possibility of metal doping of hybrid halide perovskites. Our results provide a useful theoretical guidance to rationalize and improve current doping strategies of hybrid halide perovskites with metal ions.

## Introduction

Over the past decade, hybrid halide perovskites (HHPs) have remained on the list of the hottest innovative materials for photovoltaics and optoelectronics due to their unique set of intrinsic properties, such as a direct band gap, strong optical absorption [1], long charge carrier lifetime [2,3] and remarkable tolerance to defects [4]. Despite the exceptional progress in the efficiency of solar cells and other perovskite-based devices achieved mainly due to the improvement of existing synthetic methods and optimization of device architectures [5], the potential of these strategies for further boosting the device's efficiency has been largely exhausted. Therefore, there is a growing demand for approaches to fine-tuning the optoelectronic properties of perovskites, in particular, through doping [6,7].

Importantly, the doping approaches and mechanisms of perovskites differ drastically for hybrid perovskites from those of conventional semiconductors (Si, GaAs) due to the unique chemical and structural origin of the perovskites. Primarily, the nominal doping levels (1-10 mol%) most commonly used for HHPs [7–9] are exceptionally high compared to the conventional semiconductors. Such a high concentration of impurity ions leads potentially to two scenarios: substitution (if possible in terms of the size, electronegativity, and chemical nature of the ion) or the formation of composites containing secondary phases [10,11]. The effect of composite formation in a number of cases lead to an improvement in the functional properties of the perovskite layer, but it does not relate to the doping of the material itself, possibility and character of which is the main focus of this article.

Despite the great number of experimental efforts, the rational development of doping approaches for HHPs is hindered by the absence of unambiguous answers on basic questions, concerning the possibility of inclusion of a given ion in the structure as well as its preferable lattice position. Noteworthy, in most works, the “doped” perovskites were investigated only in terms of the resulting device’s efficiency or by experimental methods with limited sensitivity, such as XRD and XPS providing no unambiguous information about the concentration of the dopant and the mechanism of doping. Unfortunately, reliable experimental techniques on the local structure such as solid-state nuclear magnetic resonance (ss-NMR) were applied only for selected metal ions [10–13].

In this article, we provide a systematic study of matching of different metal ions to available lattice sites in the structure of three most common halide perovskites: $CsPbI_3$, $MAPbI_3$, and $FAPbI_3$. For this purpose, we calculated and analyzed the effective radii for 40 most common monovalent, divalent, and trivalent metal ions in iodine environment of different compounds valuable as dopants for halide perovskites. The proposed methodology has demonstrated a good agreement with existing experimental data and is proposed to rationalize doping strategies.

## Methods

***System of ionic radii.*** To elucidate appropriate lattice sites for a given ion, a reliable system of ionic radii is required. The direct application of traditional concepts of ionic radii for iodide compounds is fraught with considerable errors due to the larger size and the less electronegativity of iodine compared to oxygen and fluorine for ionic compounds basically used in the concept of Shannon radii and tolerance factors [14,15]. There are two ways to eliminate this inaccuracy: (1) introducing corrections to the reference radius values based on the average M-X bond length [16] or (2) direct calculation of effective ionic radii from alternative appropriate sets of structural data. The latter can be accurately and facilely calculated as the equivalent radii of spherical domain ($R_{sd}$) of Voronoi-Dirichlet Polyhedra (VDP) derived (see SI), for example, from corresponding space partitioning implemented in the TOPOSPro software package [17,18].

Being calculated for each ion from a representative set of experimentally refined crystal structures of iodide compounds (including HHPs), the $R_{sd}$-based effective ionic radii, by default, take into account the shortening of the M-I bond for various metals due to a covalent bond contribution. Moreover, $R_{sd}$ for a given atom depends only on the oxidation state, electronic configuration, and the type of neighboring atoms, being approximately constant in different structures and different coordination numbers. Therefore, the single $R_{sd}$ value can be used for different possible lattice sites of a given doping ion ($M^{n+}$) in the crystal structure to elucidate the most appropriate one. In addition, the $R_{sd}$ can be calculated with a high accuracy even for ions with a high degree of polyhedral distortion and irregular coordination, which made it ambiguous to define their ionic radii within traditional hard-sphere models. From this point of view, the effective radii of ions ($R_{sd}$) are more suitable parameter for estimation of effective sizes of doping ions and its possible locations in lead halide perovskites.

***Dopability criterions.*** In principle, three possible variants of an insertion of a foreign ion into the $ABX_3$ perovskite structure can be distinguished: the substitution of cations A or B at their own sites, or occupation of an interstitial site. So, comparison of the ionic radius of an impurity ion with the size of a given crystallographic site should be considered as a first descriminating criterion of dopability. The sizes of A and B crystallographic sites in $APbI_3$ lead iodide perovskites were quantitively estimated as the $R_{sd}$ values of $A^+$ and $Pb^{2+}$ cations averaged for different polymorphs (Fig. 2). The $R_{sd}$ values for other ions considered as possible dopants were calculated from a number of corresponding compounds with metal ions in iodine environment from Materials Project database [19].

It should be noted that the effective radius of any of the ions depends substantially on the bond ionicity (e.g. $R_{sd}$ of $Pb^{2+}$ decreases significantly in the $PbI_2$-$PbBr_2$-$PbCl_2$-$PbF_2$ series), but is almost the same in similar environment even despite of the change in the structural type (e.g. from layered $PbI_2$ to perovskite $APbI_3$, see Fig. 2, a). Therefore, we assume that the sizes of ions in the B site can be assumed to be almost the same for all hybrid lead iodide perovskites. At the same time, $R_{sd}$ of a given ion in the same environment has a small scatter even for homologous structures. Typically, this scatter is within ± 5% which was chosen as the limiting deviation of the size of the ion occupying a given position (plotted as orange regions on Figures 1-2 and S3-S8 in SI).

In the case of doping via partial Pb substitution in the HHPs structures, an equally important crystal-chemical criterion for the possibility of metal doping is a difference in electronegativity of dopant metal (M) with iodine: $\Delta\chi = \chi(I) - \chi(M)$. Therefore, the offset of $\Delta\chi$ between Pb and doping metal ($\Delta(\Delta\chi)$) should not exceed the empirical value of 0.4-0.5 $eV^{-1/2}$. In the present study, the $\Delta\chi$ values were calculated based on two electronegativity scales: the Pauling one [20] and the one recently proposed by Tantardini and Oganov [21].

The third criterion that takes into account the contribution of the polarizability and ability to be polarized when chemically bound of the cation is softness. Ions with a higher softness usually form stronger complexes with iodide anions, therefore, being in an iodine environment is more energetically favorable for them. Among the many known scales of softness, we have chosen the consensus scale published in [22] as the most correlated with the experimental properties of ions.

**Table 1.** Scales of offset in Δχ within Pauling [23] and Tantardini-Oganov [24] scales and softness**. Green highlight corresponds to Δχ and softness values appropriate for doping via partial Pb substitution.

| Metal ion | Δχ offset in Pauling scale | Δχ offset in Tantardini-Oganov scale | Softness | Metal ion | Δχ offset in Pauling scale | Δχ offset in Tantardini-Oganov scale | Softness |
|---|---|---|---|---|---|---|---|
| $Au^{+}$ | -0.21 | -0.19 | 1.82 | $Ga^{3+}$ | 0.52 | 0.19 | 0.07 |
| $Pt^{2+}$ | 0.05 | -0.36 | 1.60 | $Zn^{2+}$ | 0.68 | 0.36 | -0.09 |
| $Hg^{2+}$ | 0.33 | -0.3 | 1.16 | $Mn^{2+}$ | 0.78 | 0.42 | -0.20 |
| $Pd^{2+}$ | 0.13 | -0.08 | 1.03 | $Lu^{3+}$ | 1.06 | -0.06 | -0.30 |
| $Ag^{+}$ | 0.4 | -0.26 | 0.84 | $Eu^{3+}$ | 1.13 | 0.81 | -0.45 |
| $Cu^{+}$ | 0.43 | -0.24 | 0.65 | $Sm^{3+}$ | 1.16 | 0.72 | -0.45 |
| $Bi^{3+}$ | 0.31 | -0.07 | 0.58 | $Al^{3+}$ | 0.72 | 0.1 | -0.48 |
| $Cu^{2+}$ | 0.43 | -0.24 | 0.57 | $La^{3+}$ | 1.23 | 0.13 | -0.53 |
| $Pb^{2+}$ | 0 | 0 | 0.46 | $Rb^{+}$ | 1.51 | 0.55 | -0.69 |
| $Mo^{3+}$ | 0.17 | 0.15 | 0.38 | $K^{+}$ | 1.51 | 0.55 | -0.73 |
| $Sn^{2+}$ | 0.37 | -0.06 | 0.36 | $Ba^{2+}$ | 1.44 | 0.6 | -0.76 |
| $Ni^{2+}$ | 0.42 | 0.3 | 0.29 | $Na^{+}$ | 1.4 | 0.47 | -0.80 |
| $Ge^{2+}$ | 0.32 | -0.17 | 0.27 | $Sr^{2+}$ | 1.38 | 0.49 | -0.88 |
| $Co^{2+}$ | 0.45 | 0.28 | 0.27 | $Be^{2+}$ | 0.76 | 0.2 | -0.96 |
| $In^{3+}$ | 0.55 | 0.33 | 0.21 | $Li^{+}$ | 1.35 | 0.45 | -0.97 |
| $Cd^{2+}$ | 0.64 | 0.26 | 0.17 | $Ca^{2+}$ | 1.33 | 0.42 | -0.99 |
| $Fe^{2+}$ | 0.5 | 0.3 | 0.14 | $Mg^{2+}$ | 1.02 | 0.23 | -1.02 |

**The consensus scale published in article [22] was chosen as the scale of softness of cations.

## Results and discussions

We will consider three types of point defects resulting from foreign metal ion ($M^{n+}$) implementation in $APbI_3$ perovskite structure (electric charges and lattice positions described within Kröger–Vink notation):

1) $M_A^{(n-1)\bullet}$ ($M_A^x$, $M_A^\bullet$, and $M_A^{\bullet\bullet}$ in the case of n = 1, 2, 3);

2) $M_{Pb}^{(n-2)\bullet}$ ($M_{Pb}'$, $M_{Pb}^x$, and $M_{Pb}^\bullet$ in the case of n = 1, 2, 3);

3) $M_i^{n\bullet}$ (for ions in tetrahedral interstitial sites).

**A-site doping**

Due to the large size of cuboctahedral $A^+$ cation site, its doping with a metal ion is, in fact, exclusively represented by the case of substitution with a large monovalent cation (isovalent substitution ($M_A^x$)) since all the divalent metal cations are much smaller. According to the calculated values of $R_{sd}$, among the monovalent cations, only $Cs^+$ has a satisfactory large size to partly substitute the organic $MA^+$ and $FA^+$ cations in the structure (Figure 1a, Figures S3 and S6 in SI). The effective radii of $Rb^+$, $K^+$, and $Tl^+$ are too small to occupy the A site and simultaneously they do not fit to the B site (Figure 1a, Figures S3 and S6 in SI), which indicates that these cations can not be incorporated in the $APbI_3$ structure at all. In the case of admixing of some salts based on these ions into perovskite precursor solution, the latter should lead to separate crystallization of pure perovskite phase without "dopants" and with secondary phase(s) containing $Rb^+$, $K^+$, and $Tl^+$ cations. This conclusion fully agrees with experimentally proved by ss-NMR formation of KI and $KPbI_3$ secondary phases for $K^+$ [10] as well as $RbPbI_3$, $RbI_{1-x}Br_x$, and $Cs_{0.5}Rb_{0.5}PbI_3$ ones for $Rb^+$ [11]. When non-perovskite wide-bandgap phases of potassium and rubidium are segregated on the grain boundaries [25], they might block an interfacial non-radiative recombination as well as inhibit intralayer diffusion of iodine and organic cations in solar cells, which explains the well-known improvement of optoelectronic properties [26] and suppressed hysteresis [27,28] of resulting film and solar cells respectively.

**B-site doping**

**Monovalent cations.** The two smallest alkali cations in six-fold coordination, $Na^+$ and $Li^+$, have an admissible mean ionic size to occupy the B positions in the structure of perovskites.

However, the large offset in Δχ and in hardness for these cations compared to for $Pb^{2+}$ should prevent them from replacing lead in the $APbI_3$ perovskite. Considering the significant discrepancy in the electronegativity of both the ions according to the both scales used, it can be assumed that their dopability is limited by low concentrations. The possible $Na'_{Pb}$ ($Li'_{Pb}$) substitution generates negatively charged defects, which should be compensated by positively charged defects such as iodine vacancies ($V_I^{\bullet}$) and/or holes (h) (see eqs. in SI). This conclusion agrees well with a large increase of both ionic and electronic conductivity and pronounced p-doping of $MAPbI_3$ perovskite films and single crystals when $PbI_2$ in a precursor solution is partially replaced with NaI [29–31] and significant upward shift of the Fermi level with LiI additive [32] (Table S1).

Among the monovalent cations of coinage metals ($Cu^+$, $Ag^+$, $Au^+$), the silver has the most appropriate size, Δχ and softness to be included in the B-site. Since the average $R_{sd}$ of $Cu^+$ and $Au^+$ is only of 10% smaller than that of $Pb^{2+}$ in iodine environment and Δχ is within the limits, a moderate local lattice distortion can also allow them to replace partially lead ions. Similar to sodium, doping of perovskite by these ions should lead to p-type conductivity (if the original material is intrinsic semiconductor). This result is consistent with the decrease in conductivity by one order of magnitude due to the weakening of n-type doping of resulted $MAPbI_3$ perovskite film prepared with the addition of the CuI and AgI salts [33]. Possibilities of other variants of doping are discussed in SI.

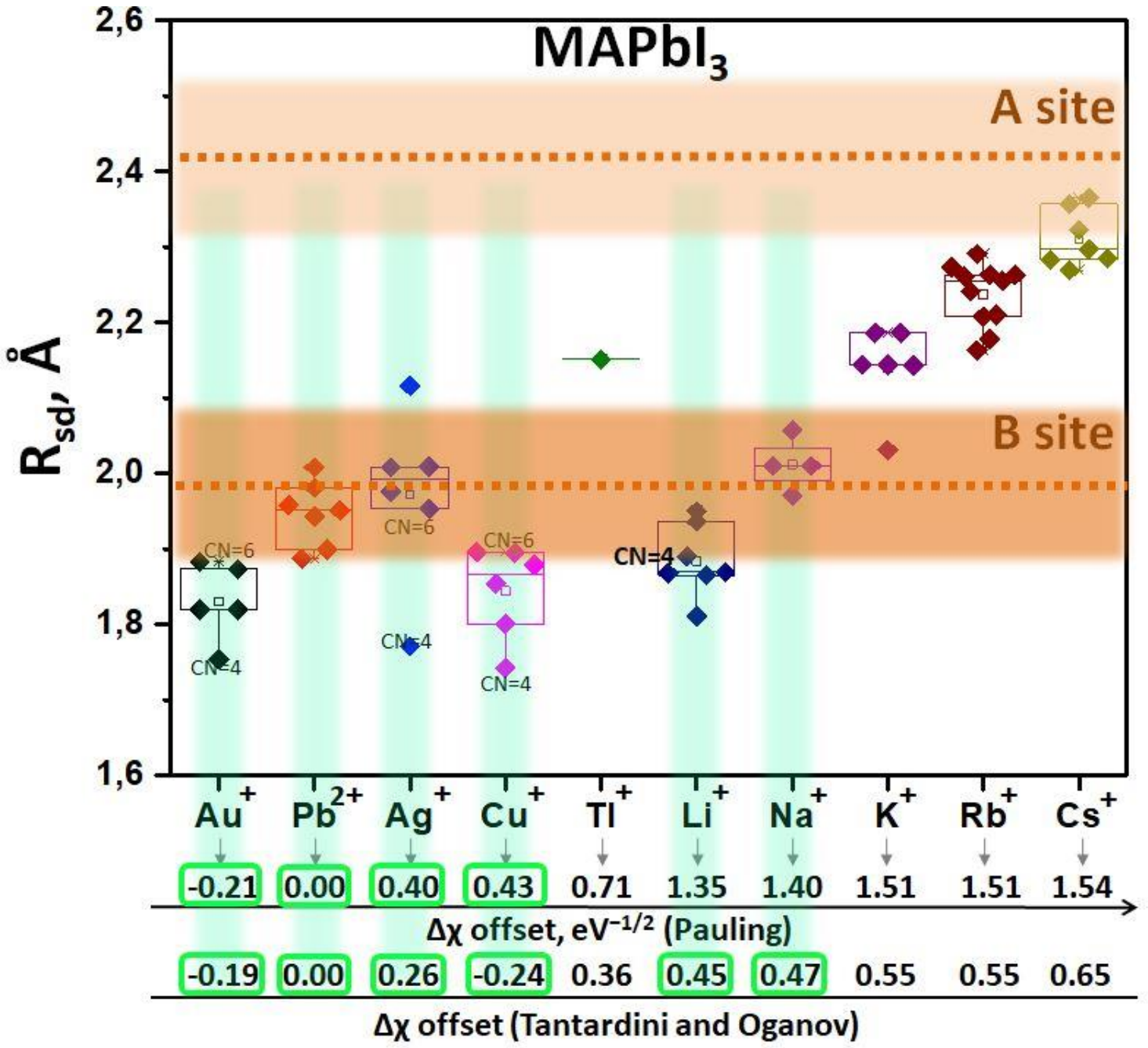
MAPbI3
A site
B site
Rsd, Å
Au+ Pb2+ Ag+ Cu+ Tl+ Li+ Na+ K+ Rb+ Cs+
-0.21 0.00 0.40 0.43 0.71 1.35 1.40 1.51 1.51 1.54
Δχ offset, eV−1/2 (Pauling)
-0.19 0.00 0.26 -0.24 0.36 0.45 0.47 0.55 0.55 0.65
Δχ offset (Tantardini and Oganov)

(a)

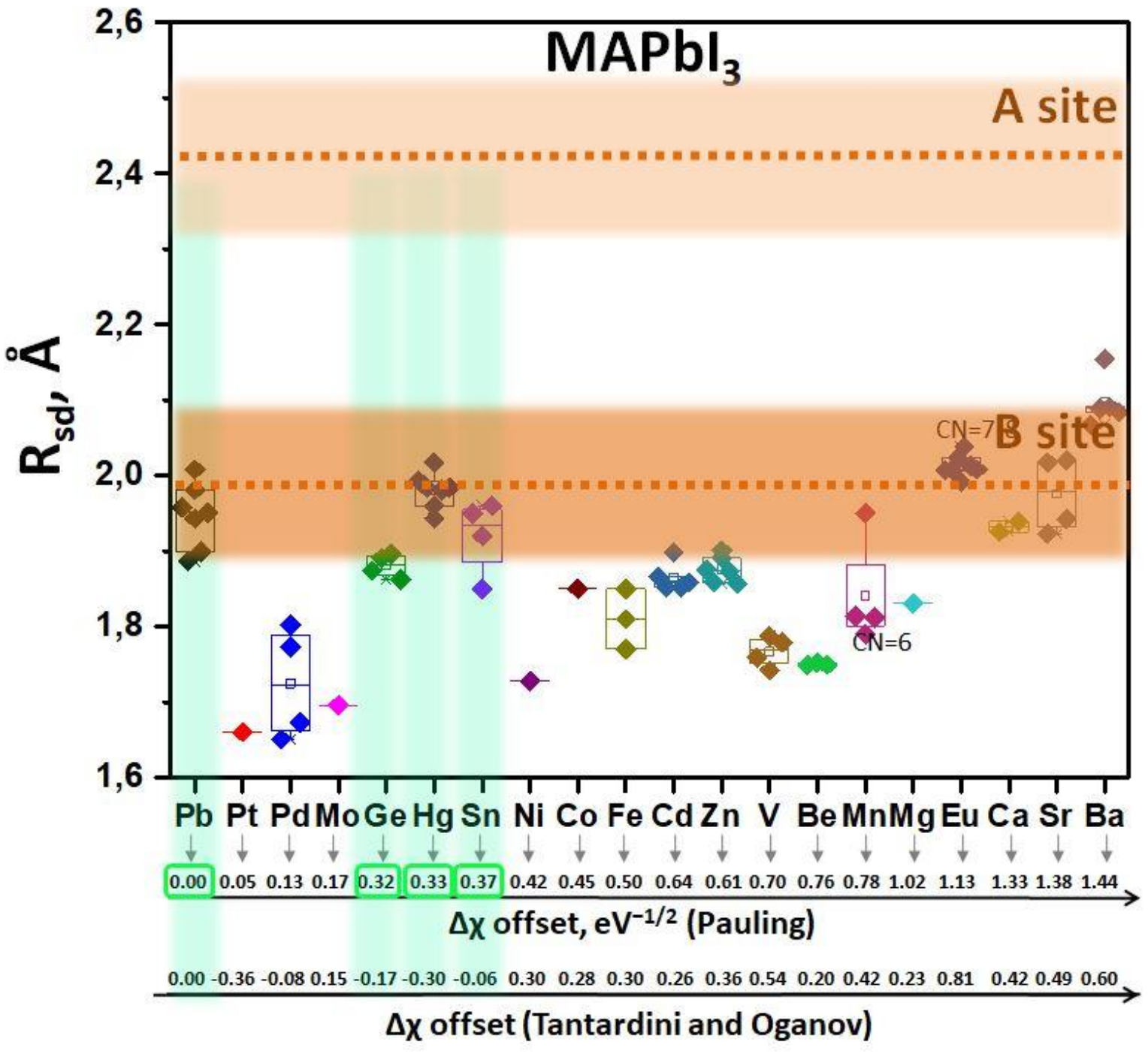
MAPbI3
A site
B site
CN=7
CN=6
Rsd, Å
Pb Pt Pd Mo Ge Hg Sn Ni Co Fe Cd Zn V Be Mn Mg Eu Ca Sr Ba
0.00 0.05 0.13 0.17 0.32 0.33 0.37 0.42 0.45 0.50 0.64 0.61 0.70 0.76 0.78 1.02 1.13 1.33 1.38 1.44
Δχ offset, eV−1/2 (Pauling)
0.00 -0.36 -0.08 0.15 -0.17 -0.30 -0.06 0.30 0.28 0.30 0.26 0.36 0.54 0.20 0.42 0.23 0.81 0.42 0.49 0.60
Δχ offset (Tantardini and Oganov)

(b)

**Figure 1.** Calculated $R_{sd}$ of monovalent (a) and divalent (b) metal ions for doping of $MAPbI_3$. The dashed lines show the values of the average $R_{sd}$ of A and B sites in $MAPbI_3$; a possible change in the sizes of A and B sites for isomorphic metal doping is shown in orange.

**Divalent cations.** For the case of doubly charged ions, due to their relatively small size, two doping scenarios can be realized: substitution of lead in B sites or occupation of a four-coordinated interstitial (tetrahedral) positions with corresponding generation of charge balancing defects. An analysis of the diagram in Fig. 2, b shows that $Sn^{2+}$ has the closest size and $\Delta\chi$ to lead in perovskite, predictably forming a continuous series of $APb_{1-x}Sn_xI_3$ solid solution in the entire range of compositions [34]. Divalent mercury is another ion that satisfies all three dopability criterions, having a nearly ideal ionic size, low $\Delta\chi$, and high softness. The solubility limit of $HgI_2$ in $PbI_2$ having been experimentally determined to be only 600 ppm [35], which can be explained by the difference in structural types of the two iodides. In the case of $MAPbI_3$, much higher doping limit can be expected from the recent study of Frolova et. al [36]. According to these data, the secondary Hg-rich phases become visible by XRD starting from 20% of $HgI_2$ and higher only.

On the other hand, there are two large groups of doubly charged cations that definitely cannot replace lead in iodide perovskites. The first of group includes $Pt^{2+}$, $Pd^{2+}$, $Mo^{2+}$, $Ni^{2+}$, $V^{2+}$, $Be^{2+}$, and $Mg^{2+}$ ions, which in similar iodide environment are much smaller than $Pb^{2+}$. Among them, the "soft" cations of d-metals such as $Pt^{2+}$, $Pd^{2+}$, $Mo^{2+}$, which prefer the four-fold iodine coordination and have a relatively small offset of $\Delta\chi$ (and, therefore, the low expected effective charge), may occupy the interstitials if they are complexed with negatively charged defects such as $V_A'$. The second group includes ions of alkaline earth metals (except the $Mg^{2+}$ and $Be^{2+}$) and $Eu^{2+}$; all of them have a large enough size to occupy the octahedral void in the perovskite structure, however the offset in $\Delta\chi$ is too large, and thus bulk doping at a reasonable level should be excluded. In the absence of suitable positions for incorporation, these ions, apparently, are not capable of doping iodide perovskites. Therefore, additives based on them should be considered exclusively as modifying (passivating) the surface or modulating the crystallization. As for the $Ba^{2+}$, this conclusion is fully corresponds to the EDS-mapping and ss-NMR results indicating respectively the non-uniform distribution and formation of secondary Ba-rich phases for $CsPb(I_{0.66}Br_{0.33})_3$ perovskite [37]. Similarly, in the case of $Sr^{2+}$, XPS and XAS spectra clearly indicate a strong enrichment of surface by the dopant and formation of surface phases [38].

Another set of divalent cations, including $Ge^{2+}$, $Co^{2+}$, $Fe^{2+}$, $Cd^{2+}$, $Zn^{2+}$, and $Mn^{2+}$, is characterized by much more ambiguous doping behavior. All these ions are smaller than lead and have a borderline size. Therefore, to differentiate their role in doping, it is necessary to compare their $\Delta\chi$ and softness. The difference in electronegativities increases in the aforementioned series from left to right, whereas softness increases in the following series $Pb^{2+} > Ge^{2+} > Cd^{2+} \approx Co^{2+} > Fe^{2+} >> Zn^{2+} > Mn^{2+}$ (Table 1), where the first four ions are principally “soft” like lead and the last two ones are relatively “hard”. Based on the two above mentioned criterions, we can assume that $Ge^{2+}$ions, which are closest to $Pb^{2+}$ in all three parameters, are most likely suitable to substitute the lead in $APbI_3$ perovskites. Interestingly, even the relatively “soft” $Cd^{2+}$ and $Co^{2+}$ cations, as confirmed by ss-NMR, can not be incorporated into the $MAPbI_3$ perovskite structure at doping levels as low as 1 mol % [12,13]. This fact can serve as an evidence of the presence of a rather sharp lower boundary of the limiting size for the ion replacing the B position at a level of about 5% of $R_{sd}$. Accordingly, for $Fe^{2+}$, $Zn^{2+}$, and $Mn^{2+}$ ions, it can be concluded that doping through $M_{Pb}^{x}$, substitution is hardly possible.

**Trivalent cations.** Among trivalent cations and rare earth elements, the $Sb^{3+}$ and $Bi^{3+}$ are the most suitable in terms of the average effective radius for entering the B site in $MAPbI_3$, $FAPbI_3$ and $CsPbI_3$ (Figure 2, Fig. S5 and S8 in SI) perovskites. It has been shown experimentally that $Bi^{3+}$ and $Sb^{3+}$ can generate $M_{Pb}^{\bullet}$ defects in halide perovskites acting as n-dopants [39,40]. Such defects also generate sub-gap levels in iodide perovskites resulting in significantly red-shifted optical bandgap [41] and acting as recombination centers lowering the photoluminescence quantum yield [42].

As for the trivalent cations of 13 group, the $In^{3+}$ have simultaneously average ionic size nearly equal to $Pb^{2+}$, very close value of softness and appropriate $\Delta\chi$. Having a lower electronegativity than lead, $In^{3+}$ is expected to create no sub-gap electronic levels due to higher splitting of $\sigma^*$ and $\pi^*$ orbitals and, therefore, can be considered as a good choice for n-doping of iodide perovskites. In contrast, $Ga^{3+}$, having average $R_{sd}$ outside of ± 5% limits and lower softness, is most likely not capable to dope the lead halide perovskites.

The trivalent cations of rare-earth metals ($La^{3+}$, $Ce^{3+}$), likely rare earth cations, have a suitable size to enter the B-sites, but a great offset in the electronegativities and differences in the softness between these cations and $Pb^{2+}$, and that should prohibit the substitution. Therefore, despite the many evidences of control the emission properties of bromide and chloride perovskite nanocrystals [43–47] by the addition of lanthanide additives, such trick are hardly possible for $APbI_3$ iodide perovskites.

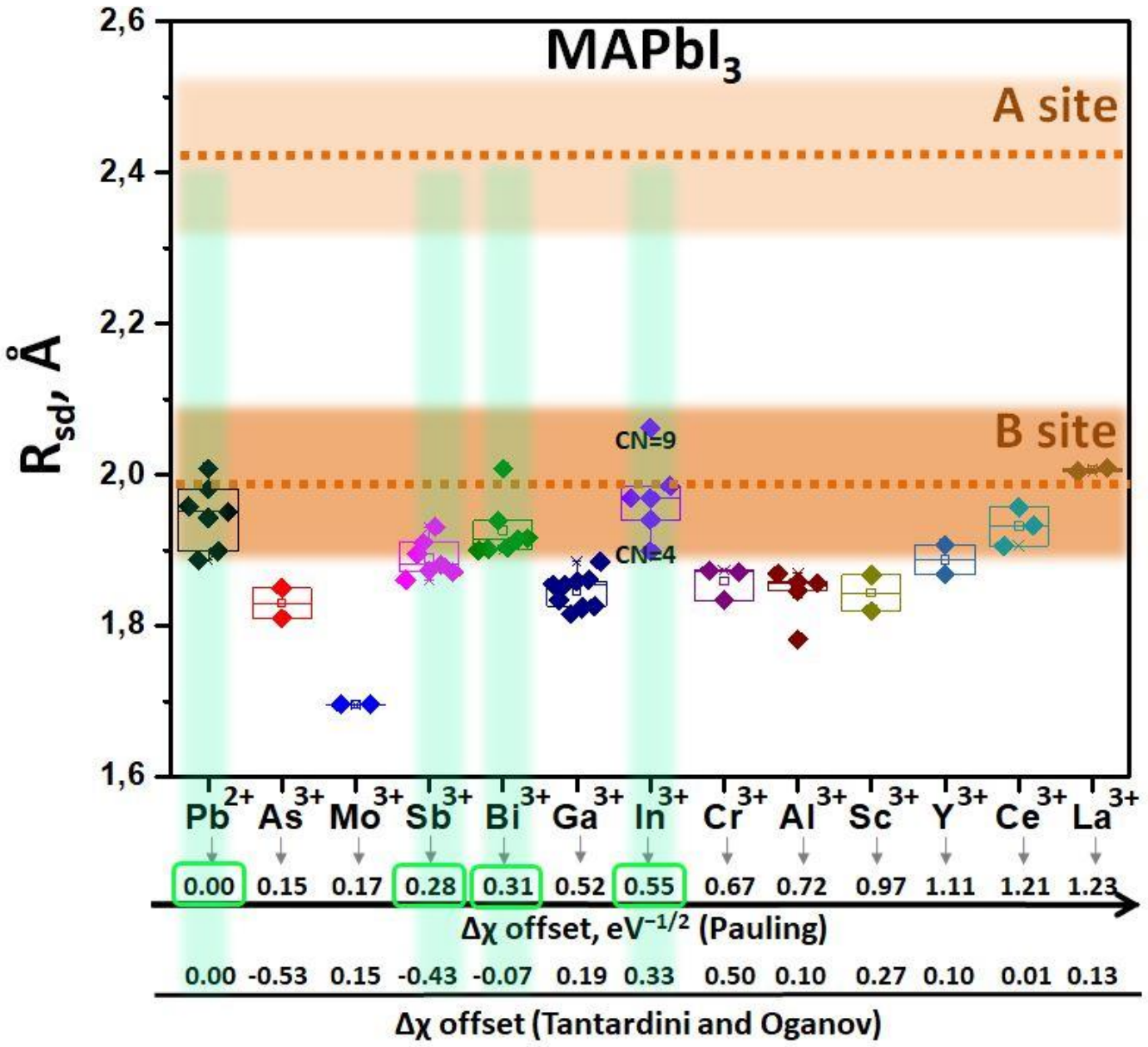

**Figure 2.** Calculated $R_{sd}$ of trivalent metal ions for doping of $MAPbI_3$. The dashed lines show the values of the average $R_{sd}$ of A and B sites in $MAPbI_3$.

**Table 2.** Appropriate lattice sites for dopability cations and their expected effect on electronic properties based on equations of point defects formation (see SI).

| Lattice site | Ions | Expected effect on electronic properties |
|---|---|---|
| A-site | $Cs^+$, $Rb^+$ | Structure (de)stabilization |
| B-site | $Ag^+$, $Cu^+$, $Au^+$ (CN = 6), $Na^+$, $Li^+$ | p-type conductivity |
| | $Sn^{2+}$, $Ge^{2+}$, $Hg^{2+}$ | Change in band structure, lowering bandgap |
| | $In^{3+}$, $Sb^{3+}$ *, $Bi^{3+}$ * | n-type doping; *sub-gap levels |
| Tetrahedral interstitial** | $Ag^+$, $Cu^+$, $Au^+$ (CN = 4) $Pt^{2+}$, $Pd^{2+}$, $Mo^{2+}$, $Ni^{2+}$ | Strong n-type doping and increased ionic conductivity due to the migration of foreign cation |
| No appropriate site | $K^+$, $Tl^+$, alkaline earth, rare earth ($M^{3+}$), divalent and | Surface passivation / Surface doping |
| | trivalent d-metal ions*** | |

**may exist in complex with negatively charged defects such as $V_A'$ or $I_A''$

*** excluded the abovementioned

The discussed considerations regarding possible dopants and their influence on the optoelectronic properties of $MAPbI_3$ are summarized in the Table 2. It should be noticed that there are obvious differences in the dopability of MA-, FA-, and Cs-based perovskites by the same ion, caused by the difference in the size of A and B lattice sites in their structures (see Figures 1-2 and Figures S3-S8). Especially, the increased $R_{sd}(Pb^{2+})$ for $FAPbI_3$ in comparison with $MAPbI_3$ makes the substitution of $M_{Pb}$ type more unlikely for ions close to the lower limit of the admissible effective radius, such as $Cu^+$, $Ag^+$, $Au^+$, and $Li^+$.

**The effect of the point defects on the doping.** All the above mentioned results are based on an assumption that the halide perovskites have perfect structure before doping. However, both the defect tolerance and soft nature of halide perovskites [48], partly aggravated by the domination of solution processing methods, often lead to the formation of various point defects at relatively high concentrations [6,49]. The requirement of bulk charge neutrality assume that there are two classic types of defects in halide perovskites: the Schottky defects (X ⇔ $V_I^\bullet$ + $V_{MA}^\bullet$ + $MA_{surf}$ + $I_{surf}$) and anionic Frenkel pairs ($I_I^x$ ⇔ $V_I^\bullet$ + $I_i'$). The presence of Schottky bivacancies may facilitate the incorporation of almost any metal ion or ionic associates due to their large size. It worth to note that the formation of interstitial iodide as a stable defect in these perovskite is hardly possible due to geometric considerations ($R_{sd}(I^-) >> R_{sd}(i_t)$), therefore, the iodide anions occupying the interstitials should be ionized. Accordingly, one should also consider an influence of such defects on the doping process.

We expect that the presence of $V_I^\bullet$ (e.g. after a prolonged annealing or in case of $PbI_2$ excess [50,51]) should increase the available sizes of $R_{sd}(A)$ and $R_{sd}(B)$. It facilitates the doping by the metal ions with $R_{sd}$ larger than $R_{sd}(Pb^{2+})$ especially in the case of heterovalent doping with $M^{3+}$ cations to the B position, associated with $I_i^x$ formation as shown by the equation 1:

A defect-free case:

$$Pb_{Pb}^x + MI_3 = M_{Pb}^\bullet + I_i' + PbI_2 \Leftrightarrow M_{Pb}^\bullet + I_i^x + e^- + PbI_2 \Leftrightarrow M_{Pb}^\bullet + e^- + PbI_2 + \tfrac{1}{2}I_2 \quad \text{(Eq. 1)}$$

A defect-influenced case:

$$Pb_{Pb}^x + V_I^\bullet + MI_3 = M_{Pb}^\bullet + I_i' + V_I^\bullet + PbI_2 \Leftrightarrow M_{Pb}^\bullet + PbI_2 \quad \text{(Eq. 2)}$$

This scenario seems to be possible for lanthanides cations and could explain the existence of Eu-doping used for well – known $Eu^{3+}$-$Eu^{2+}$ ion redox shuttle proposed to improve the stability of perovskite solar cells [52].

In contrast, the presence of $I_i'$ (e.g. formed during the photodegradation of the perovskite [53]) decreases the size of $R_{sd}$ of neighboring A and B positions and, therefore, facilitates the incorporation to the B position of cations with $R_{sd}$ smaller than $R_{sd}(Pb^{2+})$. In the case of $M^+$ cations the additional driving force would be provided by the shift of an equilibrium of the defect formation associated with the formation of $V_I^\bullet$:

A defect-free case:

$$I_I^x + Pb_{Pb}^x + MI = M_{Pb}' + V_I^\bullet + PbI_2 \quad \text{(Eq. 3)}$$

A defect-influenced case:

$$I_I^x + Pb_{Pb}^x + I_i' + MI = M_{Pb}' + V_I^\bullet + I_i' + PbI_2 \Leftrightarrow M_{Pb}' + PbI_2 \quad \text{(Eq. 4)}$$

This scenario could be realized in the case of $Li^+$ ions, migrating from Li-doped *Spiro*-MeOTAD hole transporting layer to perovskite layer upon solar cell operation [53,54].

## Conclusions

To sum up, we found that the most important parameters affecting the doping of lead iodide perovskites are the effective sizes ($R_{sd}$) of doping metal ions, their difference in electronegativity ($\Delta\chi$) and the softness. To calculate the effective sides of metal ions we introduce new scale of ionic radii based on $R_{sd}$ parameters calculated from the set of structures containing these ions in iodine environment. Using the proposed approach, we show that among monovalent alkali metal cations, $Cs^+$ and $Rb^+$ are suitable to occupy the A-site, while $K^+$ should form secondary phases and passivate the grain boundaries. The $Cu^+$, $Ag^+$, $Au^+$, $Na^+$ and $Li^+$ can occupy B-site and cause the p-doping of HHPs. The most suitable divalent cations for doping with entering in the B site are $Sn^{2+}$ and $Hg^{2+}$. Among the trivalent cations, $Sb^{3+}$, $In^{3+}$, and $Bi^{3+}$ are the most suitable to occupy the B-site and to induce the n-doping of HHPs. $Ag^+$, $Cu^+$, $Au^+$ (CN = 4), $Pt^{2+}$, $Pd^{2+}$, $Mo^{2+}$, and $Ni^{2+}$ can localize in tetrahedral interstitial positions as also associated with negatively charged defects, leading to n-type doping and increased ionic conductivity due to the migration of foreign cation. Finally, our results provide a guidance for improving current doping engineering strategies of lead iodide perovskites by metal ions.

## ASSOCIATED CONTENT

The supporting information file contains technical details of Voronoi-Dirichlet Polyhedra approach, scales of offset in $\Delta\chi$ and softness, quasi chemical equations of possible doping scenarios for monovalent, divalent and trivalent cations, comparison of calculated $R_{sd}$ of monovalent, divalent and trivalent metal ions with sizes of A- and B-site in $FAPbI_3$ and $CsPbI_3$ perovskites, doping of hybrid perovskites by metal ions and their effect on electronic properties according to literature data.

AUTHOR INFORMATION

**Notes**

The authors declare no competing financial interests.

ACKNOWLEDGMENT

This work was financial supported by a grant from RFBR (project № 19-53-53028).

REFERENCES

(1) De Wolf, S.; Holovsky, J.; Moon, S.-J.; Löper, P.; Niesen, B.; Ledinsky, M.; Haug, F.-J.; Yum, J.-H.; Ballif, C. Organometallic Halide Perovskites: Sharp Optical Absorption Edge and Its Relation to Photovoltaic Performance. *J. Phys. Chem. Lett.* **2014**, 1035–1039. https://doi.org/10.1021/jz500279b.

(2) Zhumekenov, A. A.; Saidaminov, M. I.; Haque, M. A.; Alarousu, E.; Sarmah, S. P.; Murali, B.; Dursun, I.; Miao, X.-H.; Abdelhady, A. L.; Wu, T.; et al. Formamidinium Lead Halide Perovskite Crystals with Unprecedented Long Carrier Dynamics and Diffusion Length. *ACS Energy Lett.* **2016**, 32–37. https://doi.org/10.1021/acsenergylett.6b00002.

(3) Zhumekenov, A. A. . M. I. S. O. M. B. Perovskite Single Crystals: Synthesis, Properties and Devices. In *World Scientific Handbook of Organic Optoelectronic Devices*. **2018**, 241–284.

(4) Zakutayev, A.; Caskey, C. M.; Fioretti, A. N.; Ginley, D. S.; Vidal, J.; Stevanovic, V.; Tea, E.; Lany, S. Defect Tolerant Semiconductors for Solar Energy Conversion. *J. Phys.*

*Chem. Lett.* **2014**, 1117–1125. https://doi.org/10.1021/jz5001787.

(5) Huang, J.; Yuan, Y.; Shao, Y.; Yan, Y. Understanding the Physical Properties of Hybrid Perovskites for Photovoltaic Applications. *Nat. Rev. Mater.* **2017**, 17042. https://doi.org/10.1038/natrevmats.2017.42.

(6) Zhou, Y.; Chen, J.; Bakr, O. M.; Sun, H. T. Metal-Doped Lead Halide Perovskites: Synthesis, Properties, and Optoelectronic Applications. *Chem. Mater.* **2018**, 6589–6613. https://doi.org/10.1021/acs.chemmater.8b02989.

(7) Euvrard, J.; Yan, Y.; Mitzi, D. B. Electrical Doping in Halide Perovskites. *Nat. Rev. Mater.* **2021**, 531-549. https://doi.org/10.1038/s41578-021-00286-z.

(8) Lin, Y.; Shao, Y.; Dai, J.; Li, T.; Liu, Y.; Dai, X.; Xiao, X.; Deng, Y.; Gruverman, A.; Zeng, X. C.; et al. Metallic Surface Doping of Metal Halide Perovskites. *Nat. Commun.* **2021**, 1–8. https://doi.org/10.1038/s41467-020-20110-6.

(9) Rakita, Y.; Lubomirsky, I.; Cahen, D. When Defects Become 'Dynamic': Halide Perovskites: A New Window on Materials? *Mater. Horizons.* **2019**, 1297–1305. https://doi.org/10.1039/c9mh00606k.

(10) Kubicki, D. J.; Prochowicz, D.; Hofstetter, A.; Zakeeruddin, S. M.; Grätzel, M.; Emsley, L. Phase Segregation in Potassium-Doped Lead Halide Perovskites from 39K Solid-State NMR at 21.1 T. *J. Am. Chem. Soc.* **2018**, 7232–7238. https://doi.org/10.1021/jacs.8b03191.

(11) Kubicki, D. J.; Prochowicz, D.; Hofstetter, A.; Zakeeruddin, S. M.; Grätzel, M.; Emsley, L. Phase Segregation in Cs-, Rb- and K-Doped Mixed-Cation (MA)x(FA)1-XPbI3 Hybrid

Perovskites from Solid-State NMR. *J. Am. Chem. Soc.* **2017**, 14173–14180. https://doi.org/10.1021/jacs.7b07223.

(12) Kubicki, D. J.; Prochowicz, D.; Pinon, A.; Stevanato, G.; Hofstetter, A.; Zakeeruddin, S. M.; Grätzel, M.; Emsley, L. Doping and Phase Segregation in Mn 2+ - and Co 2+ -Doped Lead Halide Perovskites from 133 Cs and 1 H NMR Relaxation Enhancement. *J. Mater. Chem. A.* **2019**, 2326–2333. https://doi.org/10.1039/c8ta11457a.

(13) Kubicki, D. J.; Prochowicz, D.; Hofstetter, A.; Walder, B. J.; Emsley, L. 113Cd Solid-State NMR at 21.1 T Reveals the Local Structure and Passivation Mechanism of Cadmium in Hybrid and All-Inorganic Halide Perovskites. *ACS Energy Lett.* **2020**, 2964–2971. https://doi.org/10.1021/acsenergylett.0c01420.

(14) Shannon, R. D. T.; Prewitt, C. T. Effective Ionic Radii in Oxides and Fluorides. *Acta Crystallogr. Sect. B Struct. Crystallogr. Cryst. Chem.* **1969**, 925–946.

(15) Shannon, R. D. Revised Effective Ionic Radii in Halides and Chalcogenides. *Acta Crystallogr.* **1976**, 751–767.

(16) Travis, W.; Glover, E. N. K. K.; Bronstein, H.; Scanlon, D. O.; Palgrave, R. G. On the Application of the Tolerance Factor to Inorganic and Hybrid Halide Perovskites: A Revised System. *Chem. Sci.* **2016**, 4548–4556. https://doi.org/10.1039/c5sc04845a.

(17) Blatov, V. A.; Shevchenko, A. P.; Proserpio, D. M. Applied Topological Analysis of Crystal Structures with the Program Package Topospro. *Cryst. Growth Des.* **2014**, 3576–3586. https://doi.org/10.1021/cg500498k.

(18) Blatov *, V. A. Voronoi–Dirichlet Polyhedra in Crystal Chemistry: Theory and

Applications. *Crystallogr. Rev.* **2004**, 249–318. https://doi.org/10.1080/08893110412331323170.

(19) Jain, A.; Ong, S. P.; Hautier, G.; Chen, W.; Richards, W. D.; Dacek, S.; Cholia, S.; Gunter, D.; Skinner, D.; Ceder, G.; et al. Commentary: The Materials Project: A Materials Genome Approach to Accelerating Materials Innovation. *APL Mater.* **2013**, 011002. https://doi.org/10.1063/1.4812323.

(20) Pauling, L. The Nature If the Chemical Bond. IV. The Energy of Single Bonds and the Relative Electronegativity of Atoms. *J. Am. Chem. Soc.* **1932**, 3570-3582.

(21) Tantardini, C.; Oganov, A. R. Thermochemical Electronegativities of the Elements. *Nat. Commun.* **2021**, 1–9. https://doi.org/10.1038/s41467-021-22429-0.

(22) Kinraide, T. B. IMPROVED SCALES FOR METAL ION SOFTNESS AND TOXICITY. *Environ. Toxicol. Chem.* **2009**, 525. https://doi.org/10.1897/08-208.1.

(23) Pauling, L. The Nature of the Chemical Bond. IV. The Energy of Single Bonds and the Relative Electronegativity of Atoms. *J. Am. Chem. Soc.* **1932**, 3570–3582. https://doi.org/10.1021/ja01348a011.

(24) Tantardini, C.; Oganov, A. R. Thermochemical Electronegativities of the Elements. *Nat. Commun.* **2021**, 1–9. https://doi.org/10.1038/s41467-021-22429-0.

(25) Kuai, L.; Wang, Y.; Zhang, Z.; Yang, Y.; Qin, Y.; Wu, T.; Li, Y.; Li, Y.; Song, T.; Gao, X.; et al. Passivating Crystal Boundaries with Potassium-Rich Phase in Organic Halide Perovskite. *Sol. RRL.* **2019**, 1–9. https://doi.org/10.1002/solr.201900053.

(26) Abdi-Jalebi, M.; Andaji-Garmaroudi, Z.; Cacovich, S.; Stavrakas, C.; Philippe, B.;

Richter, J. M.; Alsari, M.; Booker, E. P.; Hutter, E. M.; Pearson, A. J.; et al. Maximizing and Stabilizing Luminescence from Halide Perovskites with Potassium Passivation. *Nature.* **2018**, 497–501. https://doi.org/10.1038/nature25989.

(27) Tang, Z.; Bessho, T.; Awai, F.; Kinoshita, T.; Maitani, M. M.; Jono, R.; Murakami, T. N.; Wang, H.; Kubo, T.; Uchida, S.; et al. Hysteresis-Free Perovskite Solar Cells Made of Potassium-Doped Organometal Halide Perovskite. *Sci. Rep.* **2017**, 1–7. https://doi.org/10.1038/s41598-017-12436-x.

(28) Son, D.-Y.; Kim, S.-G.; Seo, J.-Y.; Lee, S.-H.; Shin, H.; Lee, D.; Park, N.-G. Universal Approach toward Hysteresis−Free Perovskite Solar Cell via Defect Engineering. *J. Am. Chem. Soc.* **2018**, jacs.7b10430. https://doi.org/10.1021/jacs.7b10430.

(29) Senocrate, A.; Moudrakovski, I.; Kim, G. Y.; Yang, T.-Y.; Gregori, G.; Grätzel, M.; Maier, J. The Nature of Ion Conduction in Methylammonium Lead Iodide: A Multimethod Approach. *Angew. Chemie Int. Ed.* **2017**, 7755–7759. https://doi.org/10.1002/anie.201701724.

(30) Yang, Y.; Zou, X.; Pei, Y.; Bai, X.; Jin, W.; Chen, D. Effect of Doping of NaI Monovalent Cation Halide on the Structural, Morphological, Optical and Optoelectronic Properties of MAPbI3 Perovskite. *J. Mater. Sci. Mater. Electron.* **2018**, 205–210. https://doi.org/10.1007/s10854-017-7905-3.

(31) Li, Y.; Li, C.; Yu, H.; Yuan, B.; Xu, F.; Wei, H.; Cao, B. Highly Conductive P-Type MAPbI3 Films and Crystals via Sodium Doping. *Front. Chem.* **2020**, 1–10. https://doi.org/10.3389/fchem.2020.00754.

(32) Fang, Z.; He, H.; Gan, L.; Li, J.; Ye, Z. Understanding the Role of Lithium Doping in Reducing Nonradiative Loss in Lead Halide Perovskites. *Adv. Sci.* **2018**, 1–6. https://doi.org/10.1002/advs.201800736.

(33) Abdi-Jalebi, M.; Pazoki, M.; Philippe, B.; Dar, M. I.; Alsari, M.; Sadhanala, A.; Divitini, G.; Imani, R.; Lilliu, S.; Kullgren, J.; et al. Dedoping of Lead Halide Perovskites Incorporating Monovalent Cations. *ACS Nano.* **2018**, 7301–7311. https://doi.org/10.1021/acsnano.8b03586.

(34) Hao, F.; Stoumpos, C. C.; Chang, R. P. H.; Kanatzidis, M. G. Anomalous Band Gap Behavior in Mixed Sn and Pb Perovskites Enables Broadening of Absorption Spectrum in Solar Cells. *J. Am. Chem. Soc.* **2014**, 8094–8099. https://doi.org/10.1021/ja5033259.

(35) Manoel, E. R.; Custódio, M. C. C.; Guimarães, F. E. G.; Bianchi, R. F.; Hernandes, A. C. Growth and Characterization of HgI2, PbI2 and PbI2:HgI2 Layered Semiconductors. *Mater. Res.* **1999**, 75–79. https://doi.org/10.1590/S1516-14391999000200006.

(36) Frolova, L. A.; Anokhin, D. V.; Gerasimov, K. L.; Dremova, N. N.; Troshin, P. A. Exploring the Effects of the Pb2+ Substitution in MAPbI3 on the Photovoltaic Performance of the Hybrid Perovskite Solar Cells. *J. Phys. Chem. Lett.* **2016**, 4353–4357. https://doi.org/10.1021/acs.jpclett.6b02122.

(37) Xiang, W.; Wang, Z.; Kubicki, D. J.; Wang, X.; Tress, W.; Luo, J.; Zhang, J.; Hofstetter, A.; Zhang, L.; Emsley, L.; et al. Ba-Induced Phase Segregation and Band Gap Reduction in Mixed-Halide Inorganic Perovskite Solar Cells. *Nat. Commun.* **2019**, 1–8. https://doi.org/10.1038/s41467-019-12678-5.

(38) Pérez-del-Rey, D.; Forgács, D.; Hutter, E. M.; Savenije, T. J.; Nordlund, D.; Schulz, P.; Berry, J. J.; Sessolo, M.; Bolink, H. J. Strontium Insertion in Methylammonium Lead Iodide: Long Charge Carrier Lifetime and High Fill-Factor Solar Cells. *Adv. Mater.* **2016**, 9839–9845. https://doi.org/10.1002/adma.201603016.

(39) Abdelhady, A. L.; Saidaminov, M. I.; Murali, B.; Adinolfi, V.; Voznyy, O.; Katsiev, K.; Alarousu, E.; Comin, R.; Dursun, I.; Sinatra, L.; et al. Heterovalent Dopant Incorporation for Bandgap and Type Engineering of Perovskite Crystals. *J. Phys. Chem. Lett.* **2016**, 295–301. https://doi.org/10.1021/acs.jpclett.5b02681.

(40) Zhang, J.; Shang, M. H.; Wang, P.; Huang, X.; Xu, J.; Hu, Z.; Zhu, Y.; Han, L. N-Type Doping and Energy States Tuning in CH3NH3Pb1-XSb2xI3 Perovskite Solar Cells. *ACS Energy Lett.* **2016**, 535–541. https://doi.org/10.1021/acsenergylett.6b00241.

(41) Wang, R.; Zhang, X.; He, J.; Ma, C.; Xu, L.; Sheng, P.; Huang, F. Bi3+-Doped CH3NH3PbI3: Red-Shifting Absorption Edge and Longer Charge Carrier Lifetime. *J. Alloys Compd.* **2017**, 555–560. https://doi.org/10.1016/j.jallcom.2016.11.125.

(42) Zhou, Y.; Yong, Z. J.; Zhang, K. C.; Liu, B. M.; Wang, Z. W.; Hou, J. S.; Fang, Y. Z.; Zhou, Y.; Sun, H. T.; Song, B. Ultrabroad Photoluminescence and Electroluminescence at New Wavelengths from Doped Organometal Halide Perovskites. *J. Phys. Chem. Lett.* **2016**, 2735–2741. https://doi.org/10.1021/acs.jpclett.6b01147.

(43) Pan, G.; Bai, X.; Yang, D.; Chen, X.; Jing, P.; Qu, S.; Zhang, L.; Zhou, D.; Zhu, J.; Xu, W.; et al. Doping Lanthanide into Perovskite Nanocrystals: Highly Improved and Expanded Optical Properties. *Nano Lett.* **2017**, 8005–8011. https://doi.org/10.1021/acs.nanolett.7b04575.

(44) Yao, J. S.; Ge, J.; Han, B. N.; Wang, K. H.; Yao, H. Bin; Yu, H. L.; Li, J. H.; Zhu, B. S.; Song, J. Z.; Chen, C.; et al. Ce3+-Doping to Modulate Photoluminescence Kinetics for Efficient CsPbBr3 Nanocrystals Based Light-Emitting Diodes. *J. Am. Chem. Soc.* **2018**, 3626–3634. https://doi.org/10.1021/jacs.7b11955.

(45) Wang, F.; Deng, R.; Wang, J.; Wang, Q.; Han, Y.; Zhu, H.; Chen, X.; Liu, X. Tuning Upconversion through Energy Migration in Core-Shell Nanoparticles. *Nat. Mater.* **2011**, 968–973. https://doi.org/10.1038/nmat3149.

(46) Zhou, B.; Tao, L.; Chai, Y.; Lau, S. P.; Zhang, Q.; Tsang, Y. H. Constructing Interfacial Energy Transfer for Photon Up- and Down-Conversion from Lanthanides in a Core–Shell Nanostructure. *Angew. Chemie - Int. Ed.* **2016**, 12356–12360. https://doi.org/10.1002/anie.201604682.

(47) An, Y. T.; Labbé, C.; Cardin, J.; Morales, M.; Gourbilleau, F. Highly Efficient Infrared Quantum Cutting in Tb3+-Yb3+ Codoped Silicon Oxynitride for Solar Cell Applications. *Adv. Opt. Mater.* **2013**, 855–862. https://doi.org/10.1002/adom.201300186.

(48) Ball, J. M.; Petrozza, A. Defects in Perovskite-Halides and Their Effects in Solar Cells. *Nat. Energy.* **2016**, 16149. https://doi.org/10.1038/nenergy.2016.149.

(49) Chen, Y.; Zhou, H. Defects Chemistry in High-Efficiency and Stable Perovskite Solar Cells. *J. Appl. Phys.* **2020**, 060903. https://doi.org/10.1063/5.0012384.

(50) Paul, G.; Chatterjee, S.; Bhunia, H.; Pal, A. J. Self-Doping in Hybrid Halide Perovskites via Precursor Stoichiometry: To Probe the Type of Conductivity through Scanning Tunneling Spectroscopy. *J. Phys. Chem. C.* **2018**, 20194–20199.

https://doi.org/10.1021/acs.jpcc.8b06968.

(51) Wang, Q.; Shao, Y.; Xie, H.; Lyu, L.; Liu, X.; Gao, Y.; Huang, J. Qualifying Composition Dependent p and n Self-Doping in CH3NH3PbI3. *Appl. Phys. Lett.* **2014**, 163508. https://doi.org/10.1063/1.4899051.

(52) Wang, L.; Zhou, H.; Hu, J.; Huang, B.; Sun, M.; Dong, B.; Zheng, G.; Huang, Y.; Chen, Y.; Li, L.; et al. A Eu 3+ -Eu 2+ Ion Redox Shuttle Imparts Operational Durability to Pb-I Perovskite Solar Cells. *Science.* **2019**, 265–270. https://doi.org/10.1126/science.aau5701.

(53) Li, Z.; Xiao, C.; Yang, Y.; Harvey, S. P.; Kim, D. H.; Christians, J. A.; Yang, M.; Schulz, P.; Nanayakkara, S. U.; Jiang, C. S.; et al. Extrinsic Ion Migration in Perovskite Solar Cells. *Energy Environ. Sci.* **2017**, 1234–1242. https://doi.org/10.1039/c7ee00358g.

(54) Ding, C.; Huang, R.; Ahläng, C.; Lin, J.; Zhang, L.; Zhang, D.; Luo, Q.; Li, F.; Österbacka, R.; Ma, C. Q. Synergetic Effects of Electrochemical Oxidation of Spiro-OMeTAD and Li+ion Migration for Improving the Performance of n-i-p Type Perovskite Solar Cells. *J. Mater. Chem. A.* **2021**, 7575–7585. https://doi.org/10.1039/d0ta12458c.

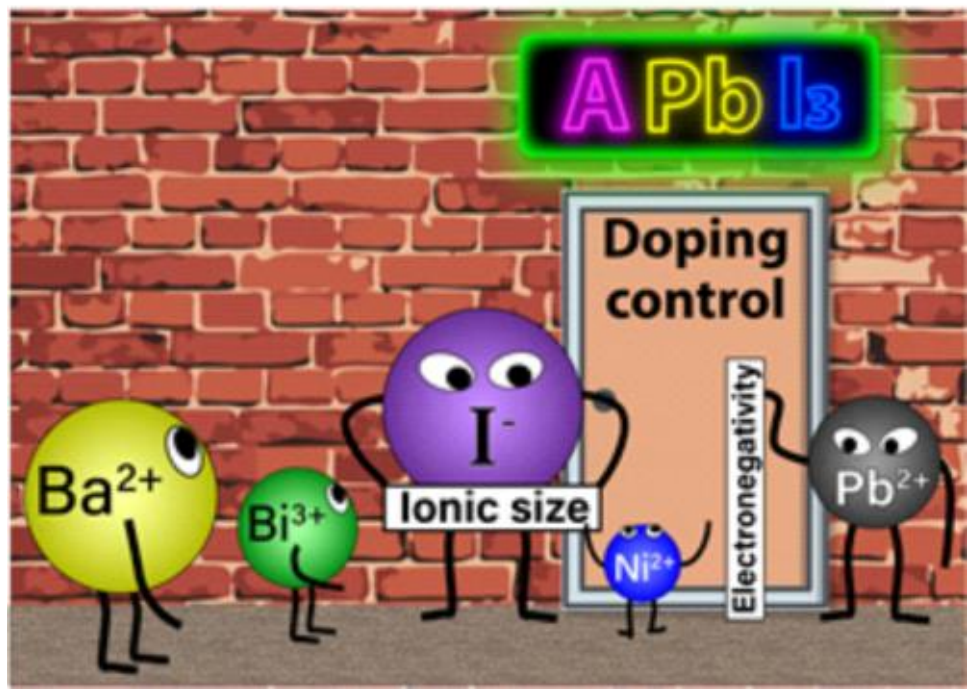

# CRYSTAL CHEMICAL INSIGHTS ON LEAD IODIDE PEROVSKITES DOPING FROM REVISED EFFECTIVE RADII OF METAL IONS

*Ekaterina I. Marchenko [1,2], Sergey A. Fateev [1], Nikolay N. Eremin [2], Chen Qi [3], Eugene A. Goodilin [1,4], and Alexey B. Tarasov [1,4]**

[1] Laboratory of New Materials for Solar Energetics, Faculty of Materials Science, Lomonosov Moscow State University, 1 Lenin Hills, 119991, Moscow, Russia

[2] Department of Geology, Lomonosov Moscow State University, 1 Lenin Hills, 119991, Moscow, Russia

[3] Beijing Key Laboratory of Nanophotonics and Ultrafine Optoelectronic Systems, School of Materials Science & Engineering, Beijing Institute of Technology, Beijing 100081, P. R. China

[4] Department of Chemistry, Lomonosov Moscow State University, 1 Lenin Hills, 119991, Moscow, Russia

Corresponding Author: * Alexey B. Tarasov, e-mail: alexey.bor.tarasov@yandex.ru

# SUPPORTING INFORMATION

**Voronoi-Dirichlet Polyhedra approach**

Effective atomic size in crystal structure correlates with geometrical parameters of atomic VDP: a convex polyhedron whose faces are perpendicular to segments connecting the central atom of VDP and other atoms. In the general case, the coordination polyhedron of an ion in crystal structure is dual to the VDP (Figure S1). The VDP characterize either relative, or the absolute atomic size in crystal structure [1].

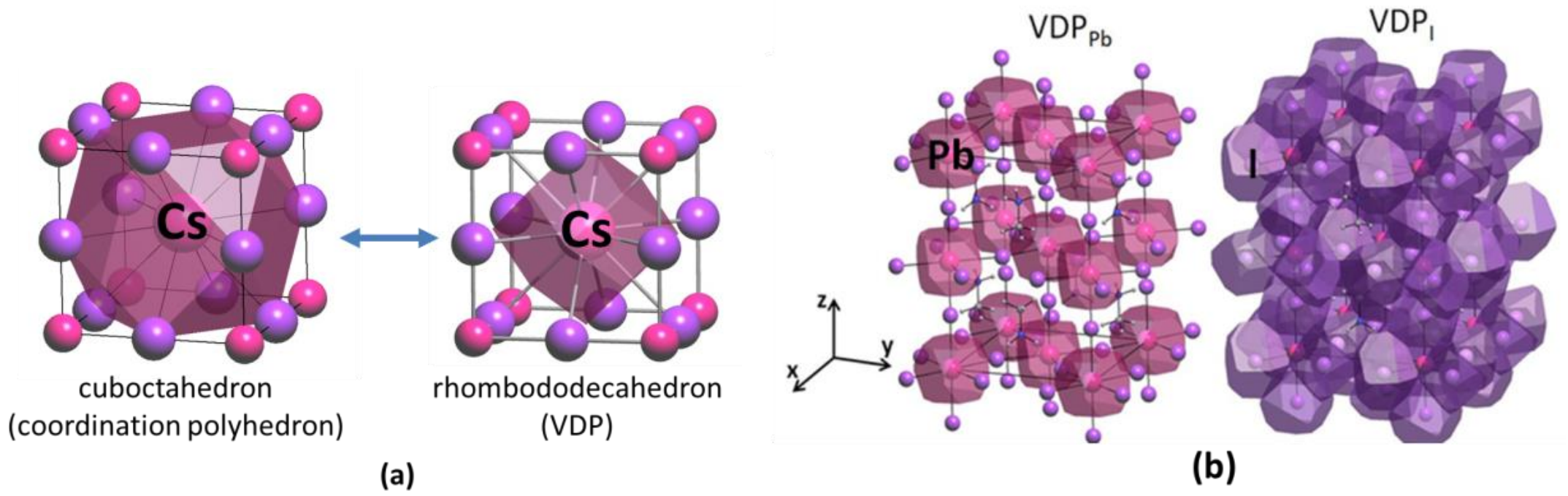

**Figure S1.** a) Coordination polyhedron (cuboctahedron) of Cs in a cubic $CsPbI_3$ perovskite and its dual VDP (rhombododecahedron); VDP for lead (left), iodine (rigth) in a tetragonal $MAPbI_3$ structure.

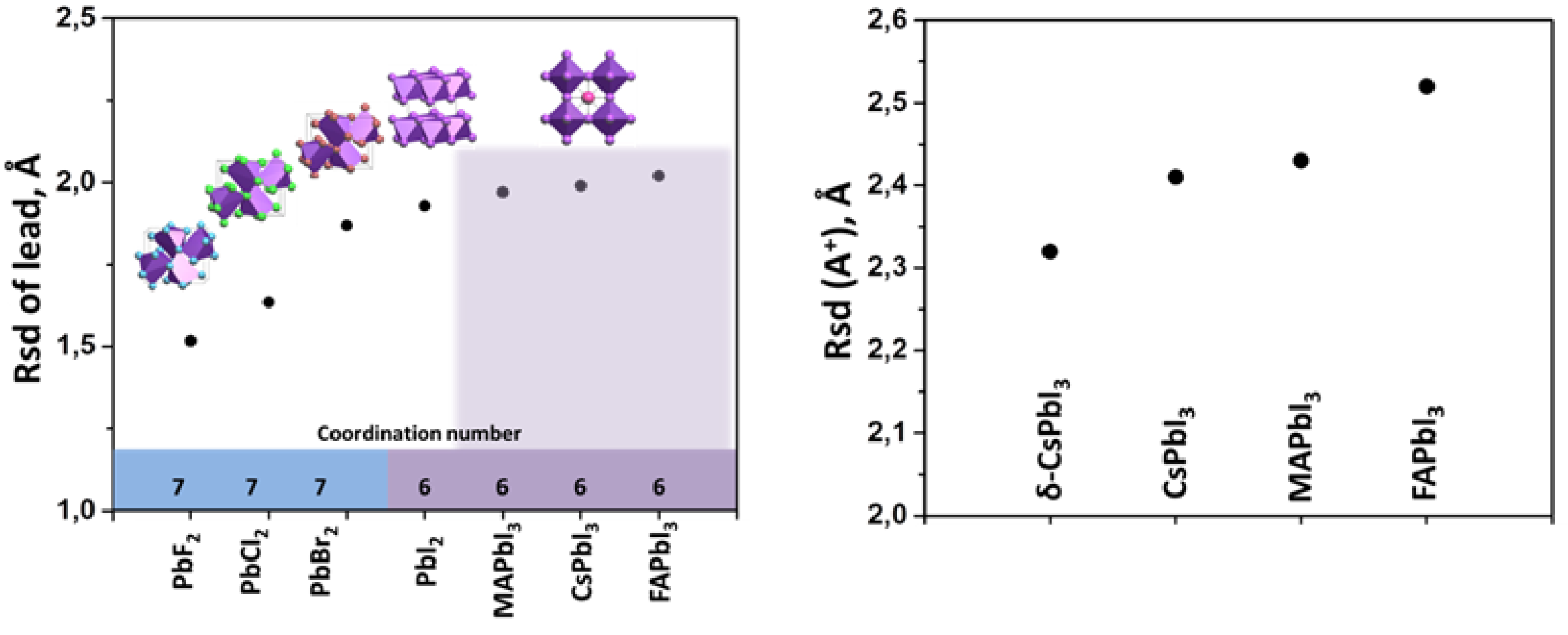

**(a)** **(b)**

**Figure S2.** The effective radii ($R_{sd}$) of (a) $Pb^{2+}$ and (b) $A^+$ cations in different crystal structures.

Comparison of calculated values of $R_{sd}$ by the Voronoi-Dirichlet partition and the Hirschfeld method for $Cs^+$ , $Pb^{2+}$ and $I^-$ in different crystal structures is shown in the Figures S9-S11.

**Possible doping scenarios for monovalent, divalent and trivalent cations**

The following are the main doping scenarios that are possible in the cases of occupation of position A by a single-charged (1), double-charged (2), and triple-charged (3) cation; occupation of position B by a single-charged (4), double-charged (5), and triple-charged (6) cation, as well as occupation of tetrahedral interstitial positions ($i_t$) by single-charged (7), double-charged (8), and triple-charged (9) cations. For each of the doping equations, possible equilibria leading to generation of charge carriers (free electrons $e^-$ and holes $h^+$) through the ionization of formed defects or the removal of components are given.

Also, each of the equations is combined with equation (0) of the self-doping of perovskite due to the formation of the most common anti-Frenkel defects.

0) $I_I^x \Leftrightarrow V_I^\bullet + I_i' \Leftrightarrow V_I^\bullet + I_i^x + e^-$ (Eq. 0)

Hereinafter, we assume that the formation of interstitial iodide as a stable defect is hardly possible due to geometric considerations (Rsd($I^-$) >> Rsd($i_t$)); therefore, the iodide anions occupying the interstitials should be ionized.

1) $M^+ \rightarrow A$

$0 + MI = M_A^x + AI$ (Eq. 1)

Scenario is possible for following ions: $Cs^+$.

Result: no effect on type of conductivity, no traps.

2) $M^{2+} \rightarrow A$

$0 + MI_2 = M_A^\bullet + I_i' + AI \Leftrightarrow M_A^\bullet + I_i^x + e^- + AI \Leftrightarrow M_A^\bullet + \mathbf{e^-} + AI + \frac{1}{2}I_2$ (Eq. 2)

Scenario is possible for following ions: ∅ (no double-charged metal ions with size enough to occupy the A site are known)

Result: n-doping

3) $M^{3+} \rightarrow A$

$0 + MI_3 = M_A^{\bullet\bullet} + 2I_i' + AI \Leftrightarrow M_A^{\bullet} + 2I_i^x + e^- + AI \Leftrightarrow M_A^{\bullet} + \mathbf{e^-} + AI + I_2$ (Eq. 3)

Scenario is possible for following ions: ∅

Result: n-doping

4) $M^{+} \rightarrow B$

$I_I^x + Pb_{Pb}^x + MI = M_{Pb}' + V_I^{\bullet} + PbI_2$ (Eq. 4.1)

$0 + MI + AI = M_{Pb}' + A_A^x + 2I_I^x + V_I^{\bullet}$ (Eq. 4.2)

$\{V_I^{\bullet} + I_i^x + e^-\} + Pb_{Pb}^x + MI \Leftrightarrow M_{Pb}' + V_I^{\bullet} + PbI_2$ (Eq. 4.3)

$\{V_I^{\bullet} + I_i^x + e^-\} + MI + AI = M_{Pb}' + A_A^x + 3I_I^x + \mathbf{h^+}$ (Eq. 4.4)

Scenario is possible for following ions: $Cu^+$, $Ag^+$, $Au^+$, $Na^+$, and $Li^+$ (if CN = 6).

Result: "dedoping" of initially n-doped perovskite, increase of ionic conductivity (due to increase of $V_I^{\bullet}$ concentration) or p-doping in the case of $[Pb^{2+}]/[M^+]$ stoichiometric replacement in solution; traps formation is possible if the ion can change the oxidation state ($Cu^+$, $Ag^+$, $Au^+$).

5) $M^{2+} \rightarrow B$

$Pb_{Pb}^{x} + MI_2 = M_{Pb}^{x} + PbI_2$ (Eq. 5.1)

$0 + AI + MI_2 = M_{Pb}^{x} + A_{A}^{x} + 3I_{I}^{x}$ (Eq. 5.2)

Scenario is possible for following ions: $Sn^{2+}$, $Hg^{2+}$, $Ge^{2+}$.

Result: no-doping, trap formation (especially, change of the oxidation state of $Sn^{2+}$ and $Ge^{2+}$).

6) $M^{3+} \rightarrow B$

$Pb_{Pb}^{x} + MI_3 = M_{Pb}^{\bullet} + I_{i}' + PbI_2 \Leftrightarrow M_{Pb}^{\bullet} + I_{i}^{x} + e^{-} + PbI_2 \Leftrightarrow M_{Pb}^{\bullet} + e^{-} + PbI_2 + \frac{1}{2}I_2$ (Eq. 6.1)

$0 + AI + MI_3 = M_{Pb}^{x} + A_{A}^{x} + 3I_{I}^{x} + I_{i}' \Leftrightarrow M_{Pb}^{x} + A_{A}^{x} + 3I_{I}^{x} + I_{i}^{x} + e^{-} \Leftrightarrow M_{Pb}^{x} + A_{A}^{x} + 3I_{I}^{x} + e^{-} + \frac{1}{2}I_2$ (Eq. 6.2)

Scenario is possible for following ions: $Bi^{3+}$, $Sb^{3+}$, $In^{3+}$.

Result: n-doping, trap formation for $Bi^{3+}$, $Sb^{3+}$.

7) $M^{+} \rightarrow i_t$

$0 + MI = M_{i}^{\bullet} + I_{i}' \Leftrightarrow M_{i}^{\bullet} + I_{i}^{x} + e^{-} \Leftrightarrow M_{i}^{\bullet} + e^{-} + \frac{1}{2}I_2$ (Eq. 7.1)

$0 + A_{A}^{x} + MI = M_{i}^{\bullet} + V_{A}' + AI$ (Eq. 7.2)

Scenario is possible for following ions: $Cu^{+}$, $Ag^{+}$, $Au^{+}$, and $Li^{+}$ (if CN = 4).

Result: n-doping, traps formation.

8) $M^{2+} \rightarrow i_t$

$0 + MI_2 = M_{i}^{\bullet} + 2I_{i}' \Leftrightarrow M_{i}^{\bullet} + 2I_{i}^{x} + e^{-} \Leftrightarrow M_{i}^{\bullet} + e^{-} + I_2$ (Eq. 8.1)

$0 + A_A^x + MI_2 = M_i^{\bullet\bullet} + I_i' + V_A' + AI \Leftrightarrow M_i^{\bullet\bullet} + I_A'' + AI$ (Eq. 8.2)

Scenario is possible for following ions: $Pt^{2+}$, $Pd^{2+}$, $Mo^{2+}$, and $Ni^{2+}$ (if CN = 4).

Result: n-doping, traps formation is possible if the ion can change the oxidation.

9) $M^{3+} \rightarrow i_t$

$0 + MI_3 = M_i^{\bullet\bullet\bullet} + 3I_i' \Leftrightarrow M_i^{\bullet\bullet\bullet} + 3I_i^x + e^- \Leftrightarrow M_i^{\bullet\bullet\bullet} + e^- + \frac{3}{2}I_2$ (Eq. 9.1)

$0 + A_A^x + MI_3 = M_i^{\bullet\bullet\bullet} + 2I_i' + V_A' + AI \Leftrightarrow M_i^{\bullet\bullet\bullet} + I_A'' + I_i^x + AI \Leftrightarrow M_i^{\bullet\bullet\bullet} + I_A'' + e^- + AI + \frac{1}{2}I_2$ (Eq. 9.2)

**Comparison of calculated $R_{sd}$ of monovalent, divalent and trivalent metal ions with sizes of A- and B-site in $FAPbI_3$ and $CsPbI_3$ perovskites**

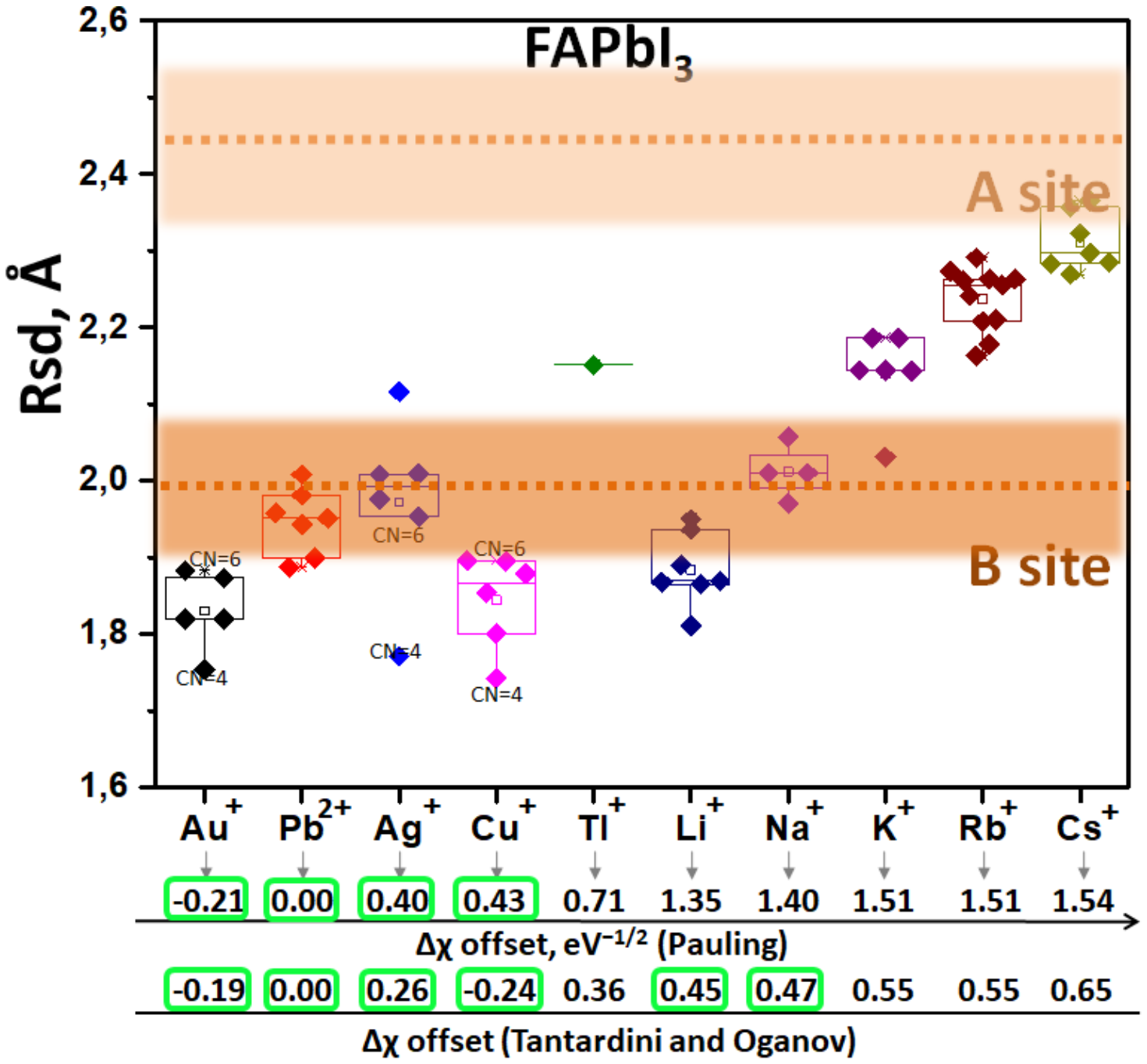

**Figure S3.** Calculated $R_{sd}$ of monovalent metal ions for doping $FAPbI_3$. The dashed lines show the values of the average $R_{sd}$ of A and B sites in $FAPbI_3$.

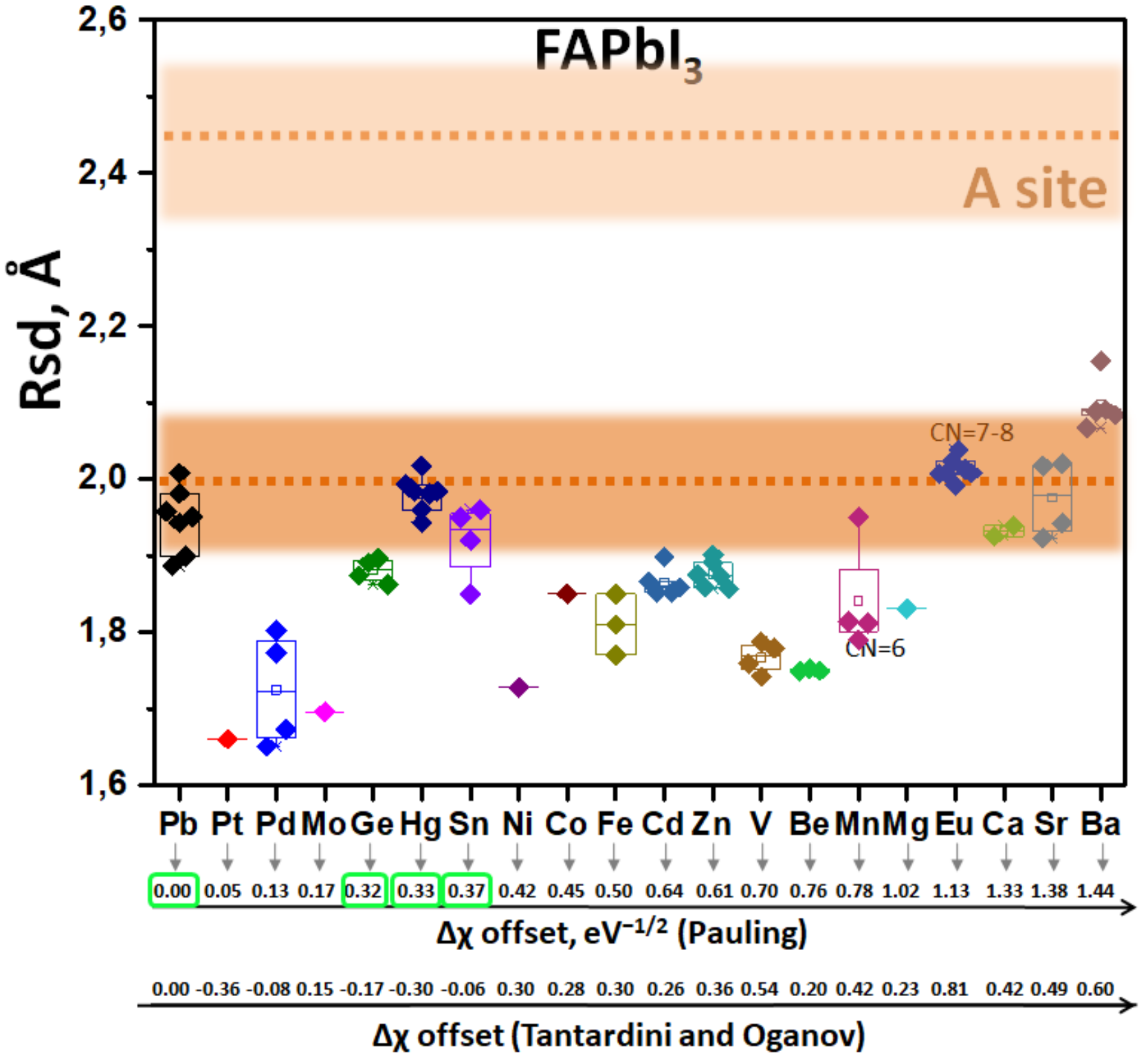

**Figure S4.** Calculated $R_{sd}$ of divalent metal ions for doping $FAPbI_3$. The dashed lines show the values of the average $R_{sd}$ of A and B sites in $FAPbI_3$.

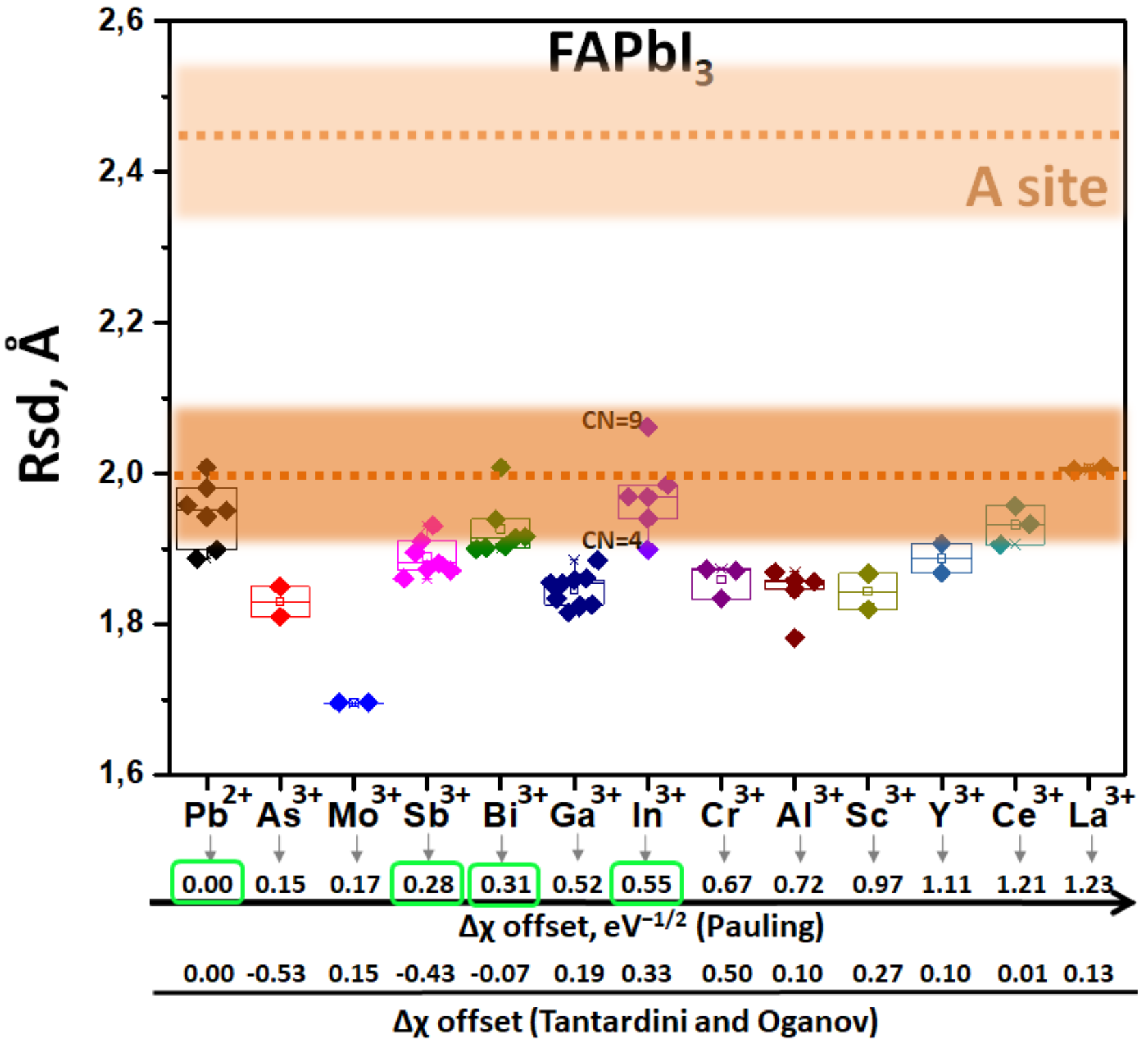

**Figure S5.** Calculated $R_{sd}$ of trivalent metal ions for doping $FAPbI_3$. The dashed lines show the values of the average $R_{sd}$ of A and B sites in $FAPbI_3$.

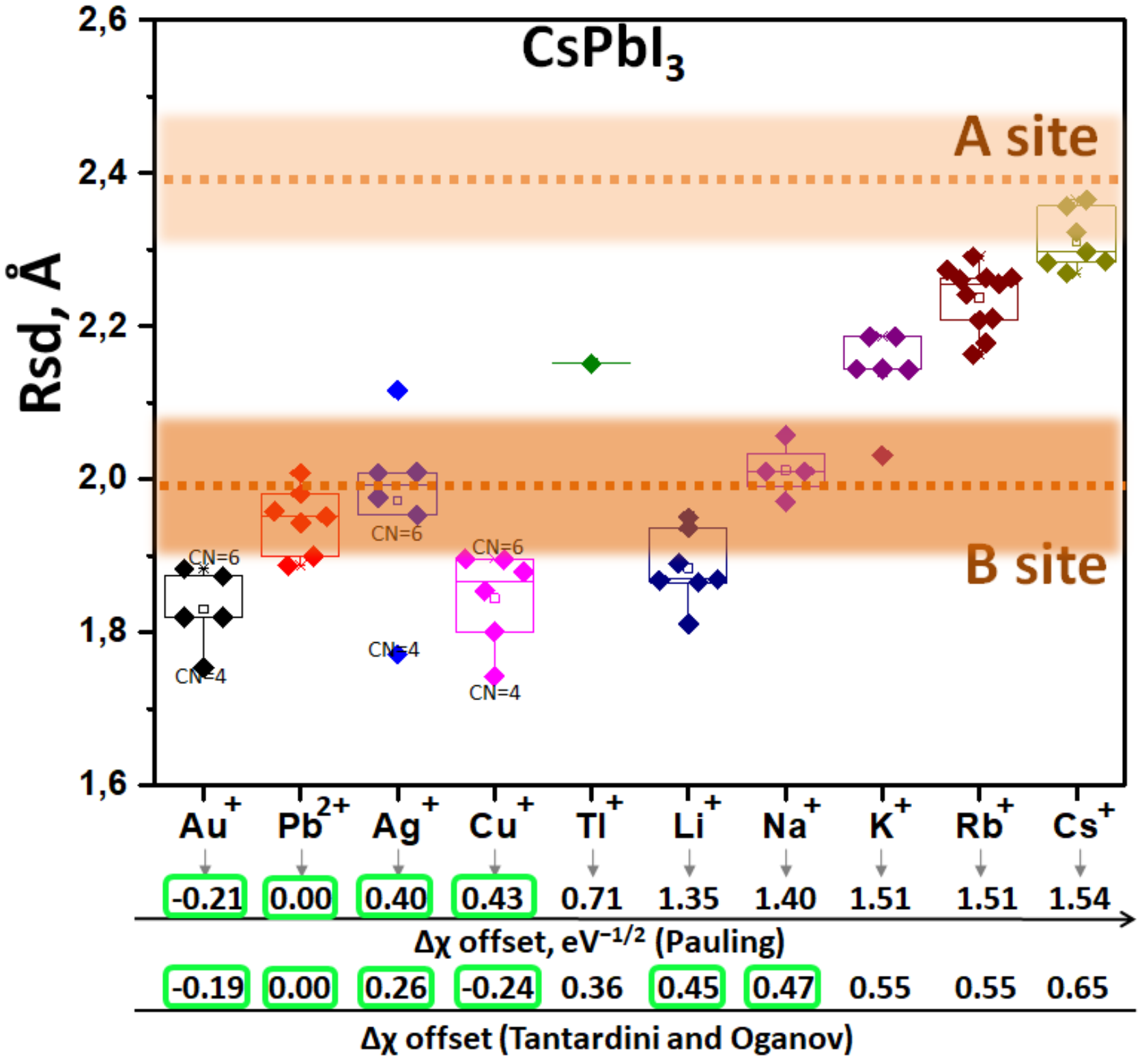

**Figure S6.** Calculated $R_{sd}$ of monovalent metal ions for doping $CsPbI_3$. The dashed lines show the values of the average $R_{sd}$ of A and B sites in $CsPbI_3$.

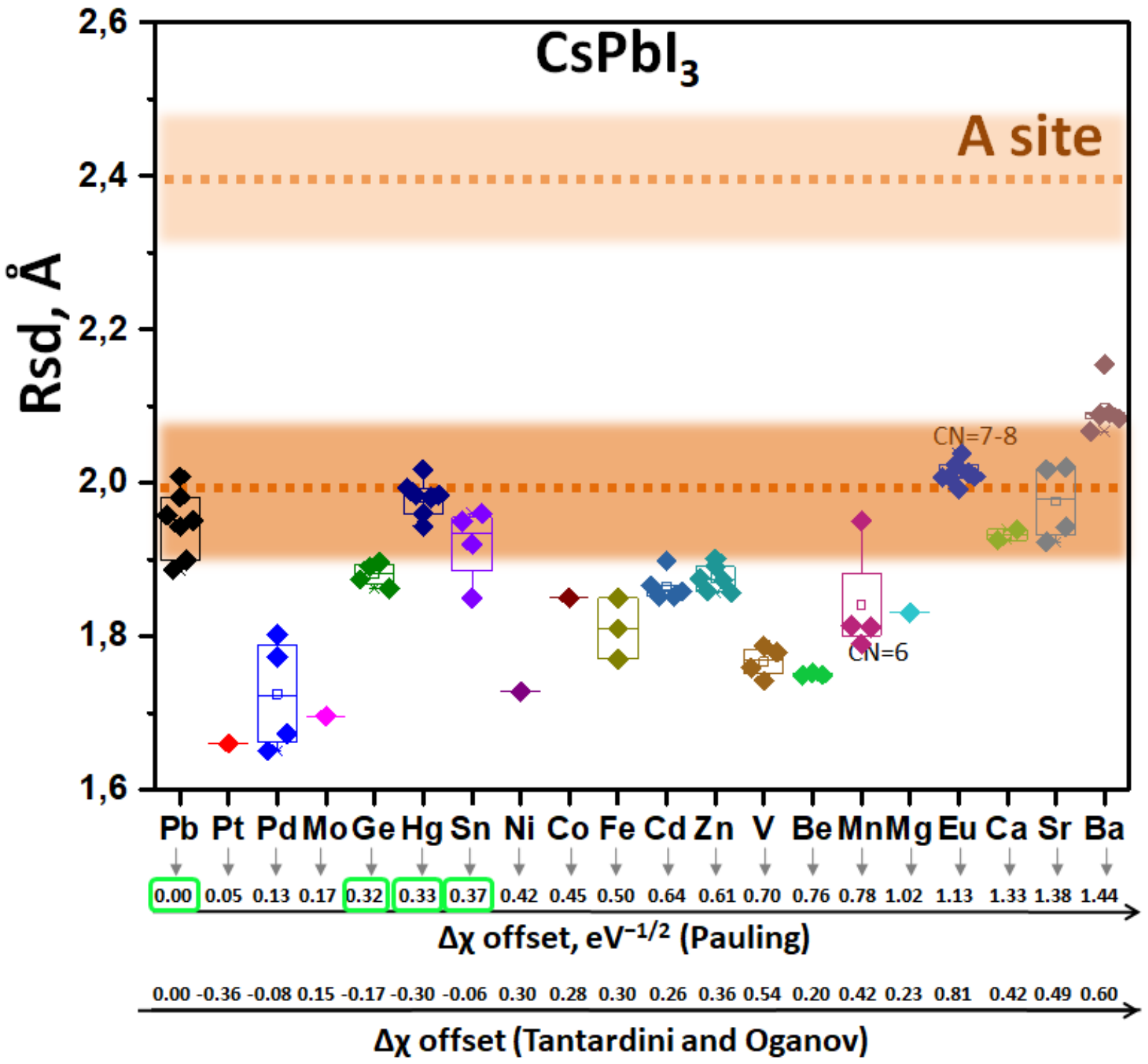

**Figure S7.** Calculated $R_{sd}$ of divalent metal ions for doping $CsPbI_3$. The dashed lines show the values of the average $R_{sd}$ of A and B sites in $CsPbI_3$.

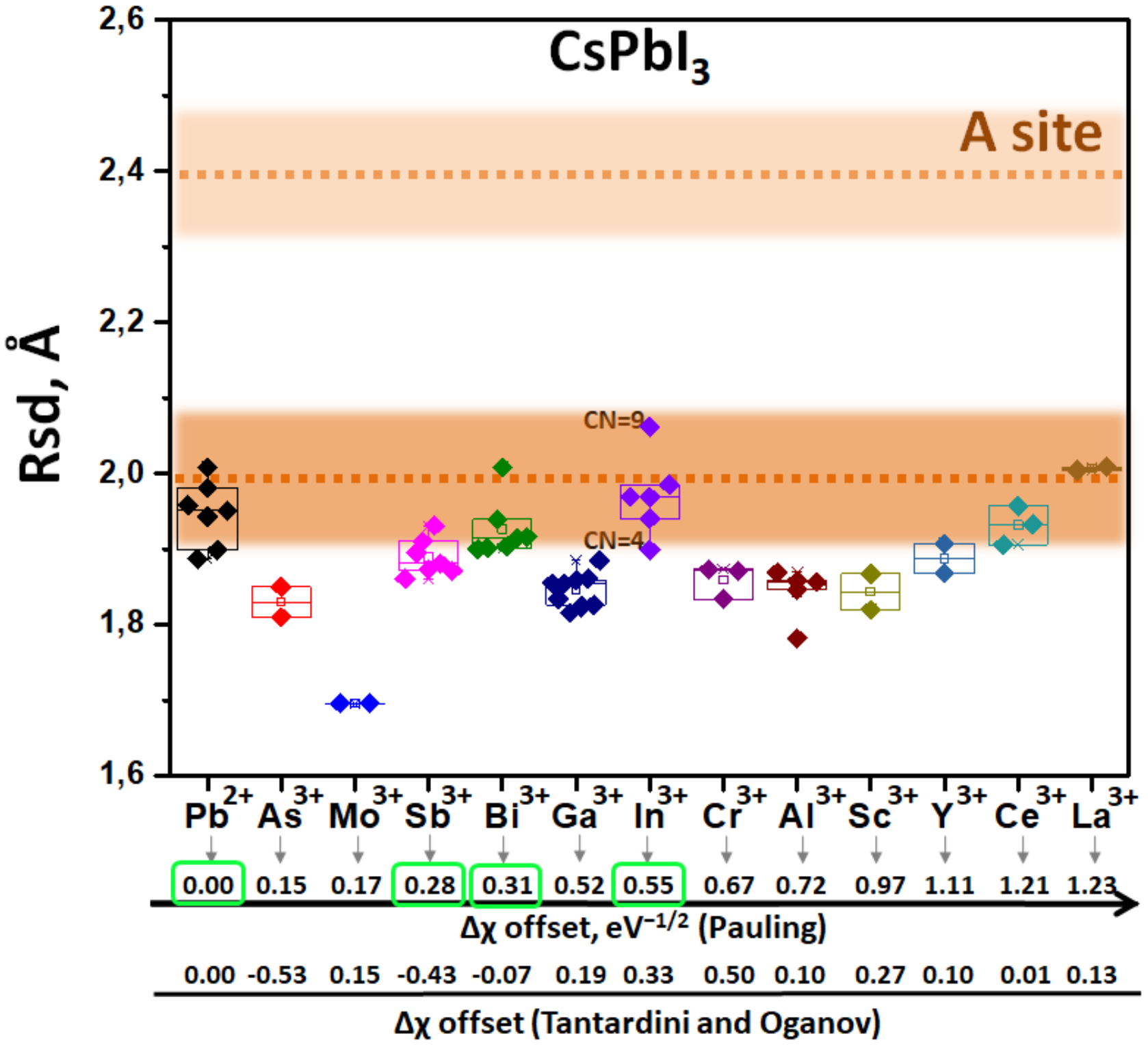

**Figure S8.** Calculated $R_{sd}$ of trivalent metal ions for doping $CsPbI_3$. The dashed lines show the values of the average $R_{sd}$ of A and B sites in $CsPbI_3$.

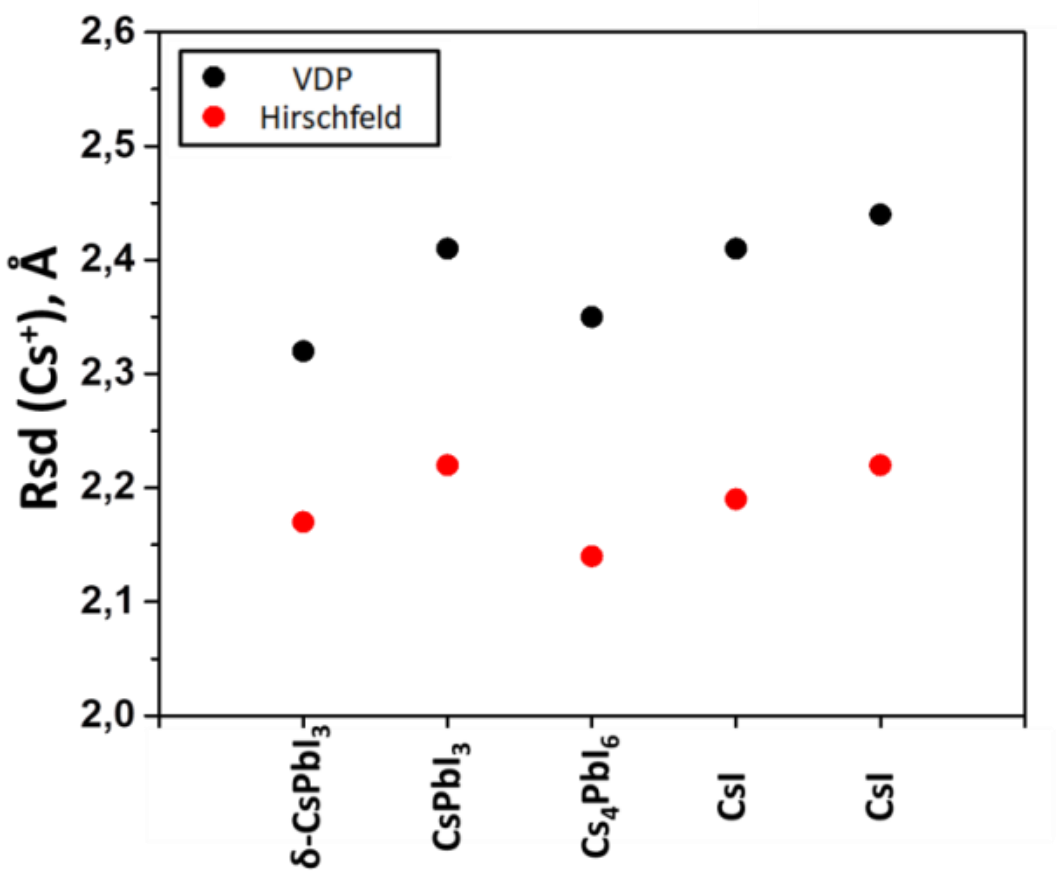

**Figure S9.** Calculated values of $R_{sd}$ by the Voronoi-Dirichlet partition and the Hirschfeld method for $Cs^+$ in different crystal structures.

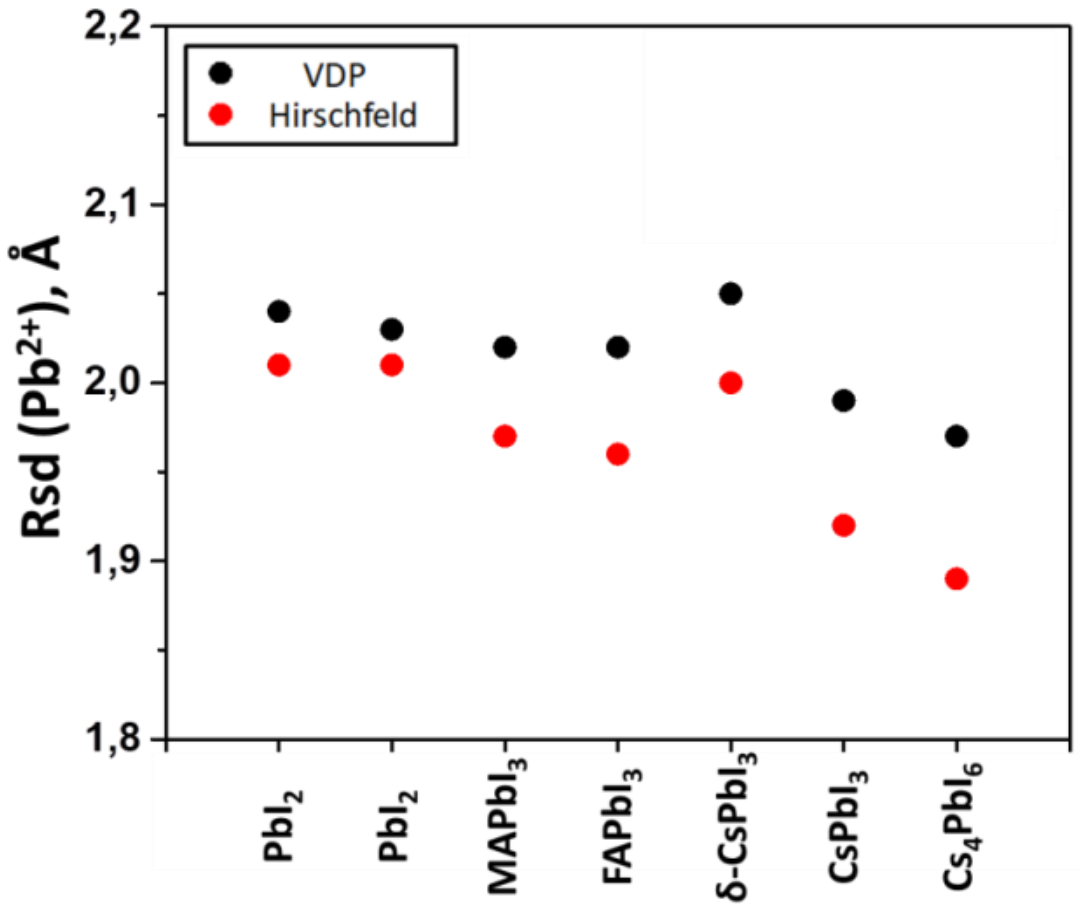

**Figure S10.** Calculated values of $R_{sd}$ by the Voronoi-Dirichlet partition and the Hirschfeld method for $Pb^{2+}$ in different crystal structures.

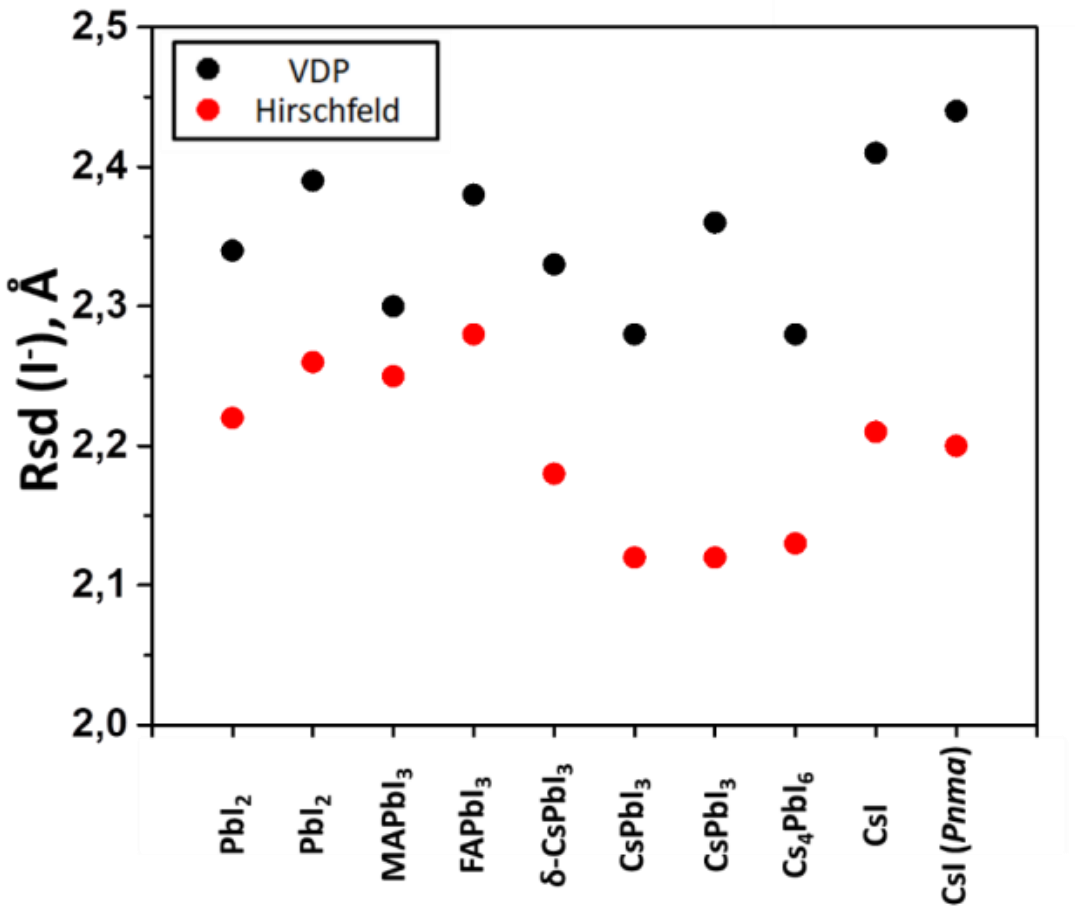

**Figure S11.** Calculated values of $R_{sd}$ by the Voronoi-Dirichlet partition and the Hirschfeld method for $I^-$ in different crystal structures.

**Table S1.** Doping of hybrid perovskites by metal ions and their effect on electronic properties according to literature data.

| Ion added to precursors | Methods | Perovskite | Result | Refs |
|---|---|---|---|---|
| $Li^+$ | PL, TRPL, UPS | $MAPbI_3$ | Work function downshift from 4.2 eV to 3.76 eV (10%); enhanced PL intensity and prolonged PL lifetime | [6] |
| $Na^+$ | HAADF STEM el. maping | $MAPbI_3$ | Formation of Na-rich phases mostly at the grain boundaries & interfaces of the perovskite film | [7] |
| $Na^+$ | PL, TPV, SIMS | $MAPbI_3$ | Enhance in PL and charge carrier lifetime for films on PTAA. Possibility of Na-doping from soda-lime glass. | [8] |
| $K^+$ | ss-NMR | | Formation of secondary phases:<br>KI, $KPbI_3$, $Cs_{0.5}K_{0.5}PbI_3$; no doping | [9] |
| $K^+$ | ss-NMR | | Formation of secondary phases:<br>KI, $KPbI_3$, $Cs_{1-x}K_xPbI_3$; no doping | [10] |
| $Rb^+$ | ss-NMR | | Formation of secondary phases:<br>$RbPbI_3$, $RbI_{1-x}Br_x$, δ-$Cs_{1-x}Rb_xPbI_3$; no doping | [10] |
| $Cs^+$ | ss-NMR | | Doping - $Cs_A^x$; solid solutions: $Cs_xFA_{1-x}PbI_3$, $Cs_xMA_{1-x}PbI_3$, $Cs_xFA_{1-x}Pb(I_{1-x}Br_x)_{3}$54 | [10] |
| $Cu^+$ | HAADF STEM el. maping, SCLC | $MAPbI_3$ | Drop in conductivity by an order of magnitude, uniform distribution | [7] |
| $Ag^+$ | HAADF STEM el. maping, SCLC | $MAPbI_3$ | Formation of Ag-rich inclusions in perovskite;<br>Drop in conductivity by an order of magnitude | [7] |
| $Ag^+$ | PL, UV-Vis spectroscopy | $CsPbI_3$ (QDs),<br>x = 1-9% | Blue shift of PL maximum < 8 nm,<br>Suppression of band tail states in abs. spectra | [11] |
| $Ag^+$ | UPS, XPS, PL, TRPL | $MAPbI_3$ | Downward shift of the Fermi level and reduced electron concentration, no shift of PL, increased carrier lifetime (TRPL) | [12] |
| | | | | |
| | | | | |
| $Mg^{2+}$ | HAXPES* | $MAPbI_3$ | Phase segregation at C($Mg^{2+}$) > 1 %<br>More n-type at ($Ca^{2+}$)<2 %, less n-type at ($Ca^{2+}$) > 2 % | [13] |
| $Ca^{2+}$ | HAXPES* | $MAPbI_3$ | Phase segregation at C($Ca^{2+}$) > 0.5 % | [13] |
| $Sr^{2+}$ | HAXPES* | $MAPbI_3$ | Phase segregation at C($Ca^{2+}$) > 0.2 %<br>More n-type at ($Ca^{2+}$)<0.5 %, less n-type at ($Ca^{2+}$)>0.5 % | [13] |
| $Ba^{2+}$ | ss-NMR,<br>EDS-mapping | $CsPbI_2Br$ | Ba/Pb = 0 – 40%: impurity barium-rich non-perovskite phases formation, non-uniform element distribution | [14] |
| $Mn^{2+}$ | ss-NMR | $CsPbBr_3$,<br>$CsPbCl_3$ | Incorporation in structure (up to at least 8mol% and 3% for $CsPbBr_3$ and $CsPbCl_3$) | [15] |
| $Co^{2+}$ | ss-NMR | $MAPbI_3$ | No doping (formation of cobalt-rich phases with) | [15] |
| $Eu^{2+}$ * | ss-NMR,<br>HAADF STEM el. maping | $CsPbI_2Br$ | Europium is incorporated into perovskite structure and uniformity distributed inside the perovskite grains. | [16] |
| $Ge^{2+}$ | XRD, PL | $CsPbI_2Br$ | Incorporation up to 20% (based on XRD data) | [17] |

*Films were annealed at 280 °C for 10 min in dry air box;

EDS – ; HAADF STEM – high-angle annular dark-field scanning transmission electron microscopy. STEM-EDX elemental analysis

PCE increase: [7] (from 16.5 to 19-20 for Na, Ag, Cu),

In the recent study it was suggested based on microstrain analysis from GIXRD data that magnesium and strontium can be incorporated into perovskite structure in form of $Mg_{Pb}^{x}$ and $Sr_{Pb}^{x}$ with maximum concentration of 1% and 0.2% respectively [13]. However, the results of hard x-ray photoelectron spectroscopy (HAXPES) suggests that no doping bulk occurs but surface doping is possible.

## REFERENCES

[1] Blatov *, V. A. *Voronoi–dirichlet polyhedra in crystal chemistry: theory and applications* // **Crystallography Reviews**, 2004, V. 10, № 4, P. 249–318, DOI: 10.1080/08893110412331323170

[2] Pauling, L. *The nature if the chemical bond. IV. The energy of single bonds and the relative electronegativity of atoms* // **J. Am. Chem. Soc.**, 1932, V. 54, № 9, P. 3570

[3] Tantardini, C., Oganov, A. R. *Thermochemical electronegativities of the elements* // **Nature Communications**, 2021, V. 12, № 1, P. 1–9, DOI: 10.1038/s41467-021-22429-0

[4] Kinraide, T. B. *IMPROVED SCALES FOR METAL ION SOFTNESS AND TOXICITY* // **Environmental Toxicology and Chemistry**, 2009, V. 28, № 3, P. 525, DOI: 10.1897/08-208.1

[5] Lin, Y., Shao, Y., Dai, J., Li, T., Liu, Y., Dai, X., *et al. Metallic surface doping of metal halide perovskites* // **Nature Communications**, 2021, V. 12, № 1, P. 1–8, DOI: 10.1038/s41467-020-20110-6

[6] Fang, Z., He, H., Gan, L., Li, J., Ye, Z. *Understanding the Role of Lithium Doping in Reducing Nonradiative Loss in Lead Halide Perovskites* // **Advanced Science**, 2018, V. 5, № 12, P. 1–6, DOI: 10.1002/advs.201800736

[7] Abdi-Jalebi, M., Pazoki, M., Philippe, B., Dar, M. I., Alsari, M., Sadhanala, A., *et al. Dedoping of Lead Halide Perovskites Incorporating Monovalent Cations* // **ACS Nano**, 2018, V. 12, № 7, P. 7301–7311, DOI: 10.1021/acsnano.8b03586

[8] Bi, C., Zheng, X., Chen, B., Wei, H., Huang, J. *Spontaneous Passivation of Hybrid Perovskite by Sodium Ions from Glass Substrates: Mysterious Enhancement of Device Efficiency Revealed* // **ACS Energy Letters**, 2017, V. 2, № 6, P. 1400–1406, DOI: 10.1021/acsenergylett.7b00356

[9] Kubicki, D. J., Prochowicz, D., Hofstetter, A., Zakeeruddin, S. M., Grätzel, M., Emsley, L. *Phase Segregation in Potassium-Doped Lead Halide Perovskites from 39K Solid-State NMR at 21.1 T* // **Journal of the American Chemical Society**, 2018, V. 140, № 23, P. 7232–7238, DOI: 10.1021/jacs.8b03191

[10] Kubicki, D. J., Prochowicz, D., Hofstetter, A., Zakeeruddin, S. M., Grätzel, M., Emsley, L. *Phase Segregation in Cs-, Rb- and K-Doped Mixed-Cation (MA)x(FA)1-xPbI3 Hybrid Perovskites from Solid-State NMR* // **Journal of the American Chemical Society**, 2017,

V. 139, № 40, P. 14173–14180, DOI: 10.1021/jacs.7b07223

[11] Lu, M., Zhang, X., Bai, X., Wu, H., Shen, X., Zhang, Y., *et al. Spontaneous Silver Doping and Surface Passivation of CsPbI 3 Perovskite Active Layer Enable Light-Emitting Devices with an External Quantum Efficiency of 11.2%* // **ASC Energy**, 2018, DOI: 10.1021/acsenergylett.8b00835

[12] Chen, Q., Chen, L., Ye, F., Zhao, T., Tang, F., Rajagopal, A., *et al. Ag-incorporated organic--inorganic perovskite films and planar heterojunction solar cells* // **Nano letters**, 2017, V. 17, № 5, P. 3231–3237

[13] Phung, N., Félix, R., Meggiolaro, D., Al-Ashouri, A., Sousa e Silva, G., Hartmann, C., *et al. The Doping Mechanism of Halide Perovskite Unveiled by Alkaline Earth Metals* // **Journal of the American Chemical Society**, 2020, V. 142, № 5, P. 2364–2374, DOI: 10.1021/jacs.9b11637

[14] Xiang, W., Wang, Z., Kubicki, D. J., Wang, X., Tress, W., Luo, J., *et al. Ba-induced phase segregation and band gap reduction in mixed-halide inorganic perovskite solar cells* // **Nature Communications**, 2019, V. 10, № 1, P. 1–8, DOI: 10.1038/s41467-019-12678-5

[15] Kubicki, D. J., Prochowicz, D., Pinon, A., Stevanato, G., Hofstetter, A., Zakeeruddin, S. M., *et al. Doping and phase segregation in Mn 2+ - and Co 2+ -doped lead halide perovskites from 133 Cs and 1 H NMR relaxation enhancement* // **Journal of Materials Chemistry A**, 2019, V. 7, № 5, P. 2326–2333, DOI: 10.1039/c8ta11457a

[16] Xiang, W., Wang, Z., Kubicki, D. J., Tress, W., Luo, J., Prochowicz, D., *et al. Europium-Doped CsPbI 2 Br for Stable and Highly Efficient Inorganic Perovskite Solar Cells* // **Joule**, 2019, V. 3, № 1, P. 205–214, DOI: 10.1016/j.joule.2018.10.008

[17] Yang, F., Hirotani, D., Kapil, G., Kamarudin, M. A., Ng, C. H., Zhang, Y., *et al. All-Inorganic CsPb1−xGexI2Br Perovskite with Enhanced Phase Stability and Photovoltaic Performance* // **Angewandte Chemie - International Edition**, 2018, V. 57, № 39, P. 12745–12749, DOI: 10.1002/anie.201807270